\documentclass{aa}
\usepackage[english]{babel}
\usepackage[varg]{txfonts}
\usepackage[usenames,dvipsnames,table]{xcolor}

\usepackage{epsfig}
\usepackage{gensymb}
\usepackage{graphicx,natbib}
\usepackage{amsmath,amssymb}
\usepackage{hyperref}
\usepackage{natbib}
\usepackage{tabularx}
\usepackage{makecell}

\usepackage{tikz}
\usetikzlibrary{arrows,shapes,positioning}
\tikzstyle{arrow}=[
    thick,
    ->,
    >=latex
    ]
\tikzstyle{nodebox}=[
    draw,
    fill=gray!15
    ]

\hypersetup{
    colorlinks=true,
    citecolor=blue,
    linkcolor=red,
    urlcolor=black
    }

\begin{document}

\def\nat{Nature}
\def\apj{ApJ}
\def\apjs{ApJS} 
\def\apjl{ApJ}
\def\apss{Ap. Sp. Sci.}
\def\aap{A\&A}
\def\mnras{MNRAS}
\def\jgr{JGR}
\def\zat{Zeitschrift f\"ur Astrophysik}

\def\cs{c_\mathrm{s}}
\def\kb{k_\mathrm{B}}
\def\mp{m_\mathrm{p}}
\def\lc{l_\mathrm{c}}
\def\Lc{L_\mathrm{c}}
\def\rc{r_\mathrm{c}}
\def\rd{r_\mathrm{disk}}
\def\tm{T_\mathrm{m}}
\def\ri{r_\mathrm{i}}
\def\md{\mathcal{M}_\mathrm{d}}
\def\Omk{\Omega_\mathrm{K}}
\def\rhoc{\rho_\mathrm{c}}
\def\fenh{f_\mathrm{enh}}
\def\nh{n_\mathrm{H}}
\def\tff{t_\mathrm{ff}}
\def\arcsec{\hbox{$^{\prime\prime}$}}
\def\kms{km s$^{-1}$}

\def\inred#1{\noindent{\textbf{\color{red}#1}}}

\title{The evolution of complex organic molecules during star formation}

\author{P. Marchand\inst{1}, A. Coutens\inst{1}, J.-C. Loison\inst{2}, V. Wakelam\inst{3}, A. Espagnet\inst{1}, F. Cruz-S{\'a}enz de Miera\inst{1}}
\institute{Institut de Recherche en Astrophysique et Planétologie, Université de Toulouse, CNRS, CNES, 9 av. du Colonel Roche, 31028 Toulouse Cedex 4, France \\ \email{pierre.marchand.astr@gmail.com}
\and Institut des Sciences Mol\'eculaires (ISM), CNRS, Univ. Bordeaux, 351 cours de la Lib\'eration, F-33400 Talence, France
\and Laboratoire d’astrophysique de Bordeaux, Univ. Bordeaux, CNRS, B18N, allée Geoffroy Saint-Hilaire, 33615 Pessac, France}

\authorrunning{P. Marchand}

\date{}

\abstract{Complex organic molecules (COMs) are thought to be the precursors of pre-biotic molecules and are observed in many protostellar sources. For this paper we studied the formation of COMs during star formation and their evolution in the midplane of the circumstellar disk up to the end of the Class I stage. We used the Analytical Protostellar Environment (APE) code to perform analytical simulations of star formation and the Nautilus code to model the chemical evolution. Most COMs mainly form during the collapse or in the disk, except the lightest (CH$_3$CCH, C$_3$H$_6$, CH$_3$OH, CH$_3$CHO, CH$_3$OCH$_3$, C$_2$H$_5$OH, CH$_3$CN, CH$_3$NC, C$_2$H$_3$CN, and CH$_3$SH), which are significantly inherited by the disk from the prestellar phase. Over the first 150 kyr of the disk, the abundances of several COMs in the midplane vary negligibly (e.g., CH$_3$CCH, CH$_3$OH, and CH$_3$CN), while others experience a variation of one order of magnitude  (e.g., C$_2$H$_3$CHO HOCH$_2$CHO, and CH$_3$COCH$_2$OH). Changing physical conditions also have an impact on the abundance profiles of COMs in the disk, and their inheritance. For example, increasing the temperature of the molecular cloud from 10 K to 15 K significantly promotes the formation of COMs in the prestellar phase, notably c-C$_2$H$_4$O and N-bearing species. Conversely, increasing the cloud mass from 2 M$_\odot$ to 5 M$_\odot$ only has a minor effect on the disk abundances in the early stages.}


\maketitle

\section{Introduction}

Complex organic molecules (COMs) are molecules with at least six atoms including carbon \citep{Herbst2009}. They have been observed in various environments: cold dark clouds \citep[e.g.,][]{Bacmann2012,Taquet2017}, protoplanetary disks \citep[e.g.,][]{Oberg2015}, low-mass protostellar sources \citep[e.g.,][]{Blake1987,Coutens2015,Jorgensen2016,Lefloch2018}, high-mass protostellar sources \citep[e.g.,][]{Bouscasse2024}, as well as comets and meteorites \citep{Bockelee2000,Ehrenfreund2001}. Detected COMs include hydrocarbons \citep[CH$_3$CCH,][]{Irvine1981}, O-bearing molecules \citep[CH$_3$OH,][]{Jennings1979}, N-bearing molecules \citep[CH$_3$CN,][]{Lovas1976}, and S-bearing molecules \citep[CH$_3$SH,][]{Zapata2015}. As they are the precursors of pre-biotic molecules, it is critical to characterize their formation and evolution. More particularly, their abundance in the disk directly affects the chemical composition of planets that form in the midplane.

The timescale of planet formation is still a debated question and used to be considered in Class II protoplanetary disks. However,  models and observations now both tend to point  toward possible planet formation at the Class 0 and Class I stages. This is supported by the rapidity of grain growth in disks that can reach millimeter sizes \citep{Kwon2009,Lebreuilly2023,Marchand2023} and may form planetesimals through the streaming instability as early as the Class 0 stage \citep{Cridland2022}. Observations also reveal planet forming material in Class 0/I disks \citep{Harsono2018}, and substructures, rings, and gaps in Class I disks \citep{Sheehan2018,Segura-Cox2020} that are possible signs of protoplanets. 

However, probing COMs in the midplane of protoplanetary disks is challenging. The low temperature of the embedded material in the midplane causes the COMs to be mostly frozen on dust grains. Instead, COMs are present in the gas phase mostly in less shielded regions at the edge of irradiated disk cavities \citep{Vandermarel2021,Booth2023} or  farther from the midplane, a conclusion drawn from  models \citep{Walsh2014} and from observations \citep{Oberg2015}. In addition, the optical thickness of the continuum emission in Class 0/I disks may mask the molecular emission \citep{Harsono2018,Nazari2023,Ohashi2023}. Future observations by the James Webb Space Telescope may bring more information about the composition of the disk ices, but to date no COMs have been detected yet \citep[see first attempts by][]{Bergner2024,Nazari2024}.

The observation of COMs in both dark clouds and disks raises the question of inheritance, particularly regarding the fraction of COMs in protoplanetary disks that originate from the interstellar medium (ISM), and which are formed during the collapse or in the disk itself. For example, methanol (CH$_3$OH) is primarily formed from the hydrogenation of CO in ices \citep{Fuchs2009,Booth2021}. Therefore, this process can only take place in regions where the temperature is below the sublimation temperature of CO ($< 30$ K). The reservoir of methanol in disks may therefore be inherited from the cold dark cloud phase. On the other hand, models show that some COMs can be formed in the disk via grain surface reactions \citep[e.g., methyl cyanide, CH$_3$CN,][]{Loomis2018}. Understanding where COMs are formed is crucial for tracing the chemical evolution from the ISM to protoplanetary disks, and for interpreting the chemical composition of planets.

Using a 3D smooth particle hydrodynamics (SPH) simulation, \citet{Yoneda2016} studied the molecular evolution of a collapsing cloud and young circumstellar disk. They determined that most COMs in their network were formed within the disk rather than in the envelope. The formation of COMs and their spatial distribution in an early disk was also studied by \citet{Coutens2020} using 3D magnetohydrodynamics (MHD) simulations. They found that the initial chemical abundances in a prestellar core may critically affect the abundance of several species (e.g., CH$_3$COCH$_3$ and NH$_2$CHO) in the early circumstellar disk. They also determined that CH$_3$CN, CH$_3$OH, CH$_3$SH, and NH$_2$CHO were inherited by their young disk, while CH$_3$CCH, CH$_3$CHO, CH$_3$OCH$_3$, CH$_3$COCH$_3$, and HCOOCH$_3$ were primarily formed during the collapse. However, prohibitive numerical costs prevent such star formation hydrodynamics simulations from following the collapse of a prestellar cloud and the subsequent evolution of a disk beyond the Class 0 phase. To address this issue, several studies employ analytical methods that require much fewer computational resources. For example, \citet{Drozdovskaya2014} and \citet{Drozdovskaya2016} used analytical models of protostellar collapse to study the evolution of methanol (CH$_3$OH) and ices in Class II protoplanetary disks.

For this work, our aim was to characterize the formation and evolution of COMs during the first stages of low-mass star formation, from the prestellar core to the end of the Class I phase. We focused on 26 COMs that have been observed in protostellar sources and studied their formation, their temporal evolution, and their sensitivity to the physical conditions. To that end, we used the Analytical Protostellar Environment (APE) code \citep{Marchand2025} to simulate the evolution of a protostellar system.
We describe our methods in Sect. \ref{sec:methods}, present our results in Sect. \ref{sec:results}, and discuss them in Sect. \ref{sec:discussion}. Section \ref{sec:conclusion} is dedicated to our conclusions.

\section{Methods} \label{sec:methods}

\subsection{The APE code}

The Analytical Protostellar Environment (APE) code \citep{Marchand2025} is a simulation software designed to provide the physical conditions of star formation to chemical models and synthetic observations. The initial condition is a Bonnor-Ebert sphere \citep{Ebert1955,Bonnor1956} that collapses to form a central object, a disk, and an outflow, which are surrounded by an envelope. The code computes density and temperature maps of these protostellar environments on a 2D polar grid.   
The maps are generated from several input physical parameters, such as the mass of the cloud and the age of the central object, using analytical models and results of hydrodynamics simulations. APE easily interfaces with the publicly available codes Nautilus \citep{Ruaud2016}, RADMC-3D \citep{Dullemond2012}, and Imager.\footnote{\url{http://www.iram.fr/IRAMFR/GILDAS}\\ \url{https://imager.oasu.u-bordeaux.fr}}

We used the ``grid of particles'' mode of APE. In this mode, the code first generates a snapshot of the system at the desired time $t_\mathrm{final}$. It then places virtual particles in each cell of this map and computes their trajectory backward in time until the initial condition $t=0$ is reached. The density and temperature history of each particle can then be used as input for Nautilus, alongside initial abundances and a chemical reaction network, to compute its chemical evolution. The chemical abundance maps can finally be reconstructed into a grid from the final abundances of all particles at time $t_\mathrm{final}$. 

The APE code also offers the use of the temperature from radiative transfer simulations. To that end, maps of the grain size-distribution are generated every 1000 years after the formation of the central object. Radiative transfer Monte Carlo simulations are then run on those snapshots using RADMC-3D to compute the temperature maps of the dust. We assume that the gas and the grains are thermally coupled at the considered densities \citep[$> 10^5$~cm$^{-3}$, see][]{Merello2019} and have the same temperature. For each particle at each time step, the temperature is then interpolated in time and space from those maps.

\subsection{Physical setup}

In this study, we considered a cloud with an initial mass of 2 M$_\odot$, at times of 10, 50, 100, and 150 kyr after the formation of a central object, which occurs at $t=\tff \approx 152$ kyr. The temperature is initially uniform at $T_\mathrm{mc}=10$~K, then is calculated from the luminosity of the central object. The system at 10 kyr is our reference case. 
The disk size is determined by the mass-to-flux ratio, which we choose to be $\lambda=5$ as it is a standard value in protostellar collapse simulations \citep[e.g.,][]{Masson2016,Wurster2016,Marchand2023}. The disk scale height is influenced by the $\alpha$-disk parameter \citep{Shakura1973} that we set at $\alpha=0.01$. The initial angular velocity of the cloud is set at $\Omega_0=2\times 10^{-15}$~s$^{-1}$. We assumed a dust-to-gas mass ratio of 1\% and we enabled dust coagulation, which had an effect only on the radiative transfer calculations (not the chemistry, see Sect. \ref{sec:methods:chemistry}).
For the radiative transfer calculations, the grid was composed of 5625 cells, with a logarithmic radial sampling of 75 points between $r=1$ au and $r=1000$ au, and a uniform poloidal sampling of 75 points between $\theta=0$ and $\theta=\pi/2$. From Sect. \ref{sec:time-COM}, we focus on the midplane of the disk. The grid of particles simulations have been performed using a reduced grid with 1 au sized cells distributed linearly from the central object to the edge of the disk. At each time step, APE computes the characteristic radius of the disk, beyond which the surface density can either sharply drop or follow an exponential cut-off, depending on the user's choice. In this study, we adopted the sharp cut-off approach.

The time step in the trajectory simulations is defined as
\begin{equation}
    dt = \mathrm{min}\left(\frac{1}{f}\frac{r}{v_r}, dt_0\right), 
\end{equation}
where $dt_0=200$~yr, $r$ is the radial distance to the central object, $v_r$ the radial velocity, and $f$ a limiting factor. A small $f$ value can lead to imprecision on the density and temperature histories, which can be reflected on the final chemical abundances. It is recommended to set it to values larger than 100 \citep{Marchand2025}. Here we used $f=300$.

\subsection{Chemical model} \label{sec:methods:chemistry}

Each particle history is used as input for the Nautilus code \citep{Ruaud2016}, which performs three-phase gas-grain chemical calculations (gas phase, grain surface, grain mantle). 
The chemical network used in the Nautilus code has been widely updated recently for a better description of the COMs on grains and in the gas phase particularly some of those studied in this work such as C$_2$H$_5$OH, HOCH$_2$CHO, CH$_3$COCH$_3$, C$_2$H$_5$OCH$_3$, CH$_3$OCH$_2$OH, HOC$_2$H$_4$OH, HOCHCHCHO, CH$_3$COCH$_2$OH, CH$_3$NC, and CH$_3$NCO \citep{Manigand2021, Coutens2022, RN11591}. The network includes 828 species as well as a total of 17871 reactions. It accounts for cosmic-ray induced reactions for the gas phase H, H$_2$, He, N, N$_2$, O, and CO The UV generated by the collisions of electrons, emitted during the ionization of H$_2$ by cosmic rays \citep{Gredel1989}, are taken into account in both the gas and the solid phases. The reactions induced by X-rays and the photo-dissociation by interstellar UV photons are however not included..

The grain surface and the mantle were both chemically active for these simulations. A sticking probability of 1 was assumed for all neutral species while sublimation could occur by thermal and nonthermal (cosmic-ray induced UV photons, chemical desorption) processes including sputtering of ices by cosmic-ray collisions \citep{RN11329}. Surface reactions formalism and more detailed description of the simulations can be found in \citet{Ruaud2016}. The diffusion energies were set to be a fraction of the binding energy for the surface (0.4 times) and the mantle (0.8 times), with a diffusion barrier thickness of 0.25 nm. The binding energies are updated values using the same methodology as in \citet{Wakelam2017}, and are listed in Table \ref{tab:COMs} for the species considered in this work.

Initially, we ran Nautilus for $t_\mathrm{pre}=10^6$ years under constant physical parameters characteristic of prestellar core conditions: a density of $n_0=10^4$ cm$^{-3}$, a temperature of $T_0=10$ K, and an extinction of $A_\mathrm{v}=5$. We assumed a cosmic-ray ionization rate of $\zeta_\mathrm{CR}=1.3\times 10^{-17}$ s$^{-1}$. We started this run using the same atomic abundances as \citet{Ruaud2018}, which we detail in Table \ref{tab:ini_abundances} (with the exception of Mg, which is not included in the chemical network).
In the rest of the paper, we call ``initial abundances'' the final abundances of this run, that served as the initial conditions for the chemical simulations along each particle trajectory.
Since Nautilus does not support yet an evolving grain size-distribution \citep[see][for a discussion]{Marchand2025}, we used a unique, constant grain size that we assumed to be 0.1 $\mu$m, with a dust-to-gas mass ratio of 0.01. The bulk density of the grains is 3 g cm$^{-3}$, and their surface site density of $1.5 \times 10^{15}$ cm$^{-2}$.
In this work, we did not compute the chemical abundances in the outflow cavity, where they were set equal to the initial abundances. We considered that the sole impact of the outflow was on temperature maps, which could in turn affect chemical abundances.

\begin{table}
  \caption{Initial atomic abundances.}
  \label{tab:ini_abundances}
\centering
\begin{tabular}{llll}
\hline\hline
  Species & Abundance (/$n_{\mathrm{H}}$) & Species & Abundance (/$n_{\mathrm{H}}$)\\
\hline
 H & 1 & Si$^+$ & $1.8 \times 10^{-6}$ \\
 He & 0.09 & Fe$^+$ & $2.0 \times 10^{-7}$\\
 N & $6.2 \times 10^{-5}$ & Na$^+$ & $2.3 \times 10^{-7}$\\
 O & $3.3 \times 10^{-4}$ & P$^+$ & $7.8 \times 10^{-8}$\\
 C$^+$ & $1.8 \times 10^{-4}$ & Cl$^+$ & $3.4 \times 10^{-8}$\\
 S$^+$ & $1.5 \times 10^{-5}$ & F & $1.8 \times 10^{-8}$\\
\hline
\end{tabular} 
\end{table}

\section{Results} \label{sec:results}

\subsection{System evolution}

The resulting properties of the disk and central object in the snapshots are listed in Table \ref{tab:disk_star_prop}. By $t=\tff+150$ kyr (about 2 $\tff$), almost all the envelope mass has been accreted. At all times, the disk:central object mass ratio remains at $\sim 1:3$. The luminosity mostly stems from accretion. It increases from 29.8 L$_\odot$ to 72.8 L$_\odot$ as the central object grows in mass, which warms up the envelope. The disk grows in size from 52.6 au at $t=\tff+10$ kyr to 92.3 au at $t=\tff+150$ kyr.

\begin{figure}
    \centering
    \includegraphics[width=0.5\textwidth, trim=5cm 0cm 0cm 0cm,clip]{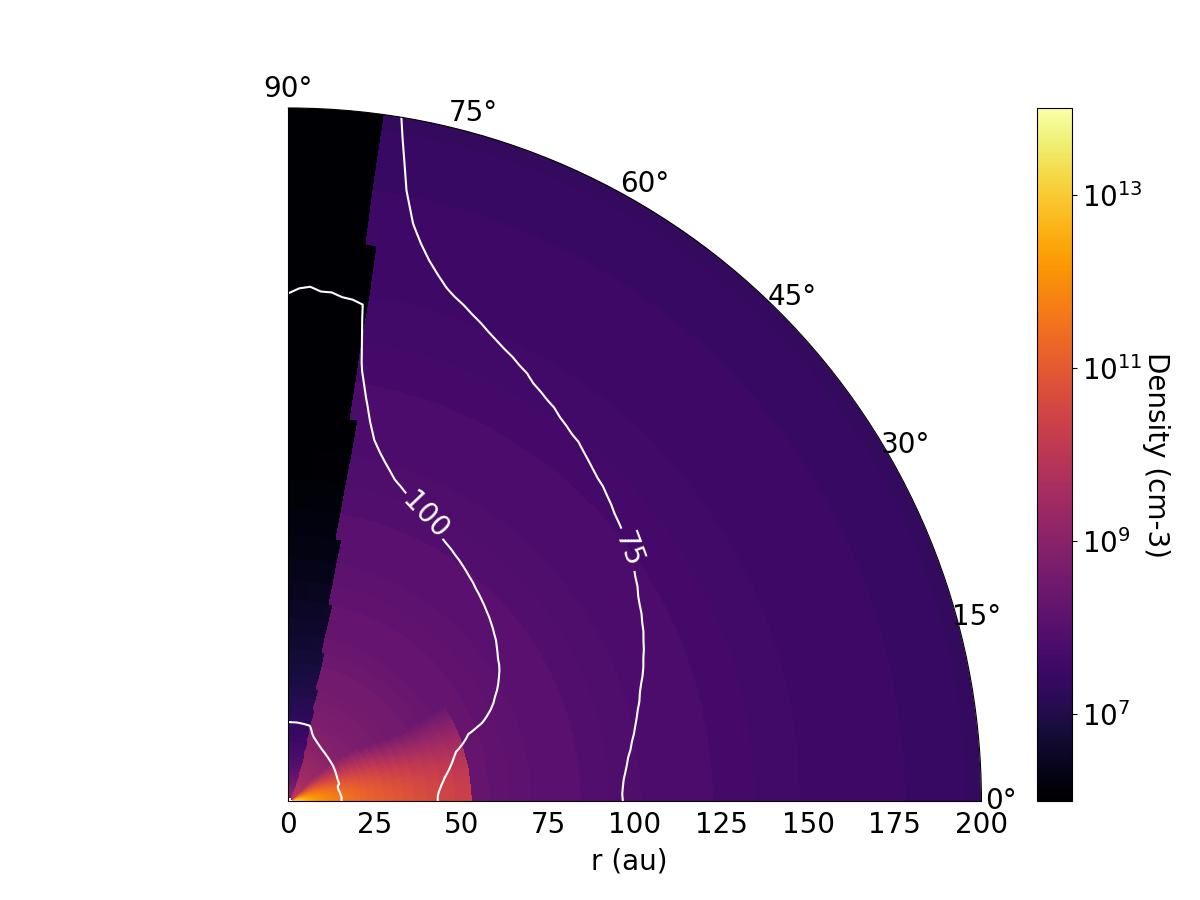}
    \includegraphics[width=0.5\textwidth, trim=5cm 0cm 0cm 0cm,clip]{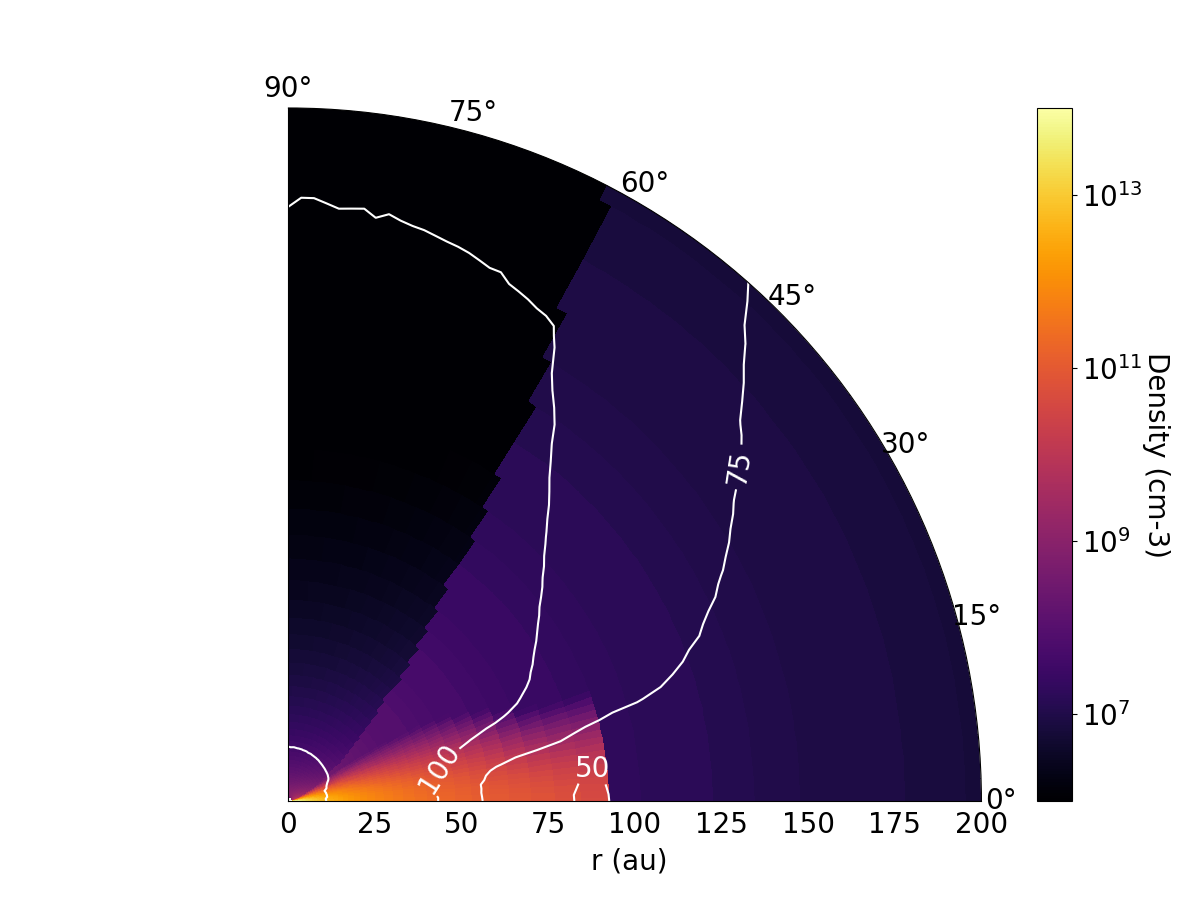}
    \caption{Zoomed-in maps of the system at $t=\tff+$10 kyr (top) and $t=\tff+150$ kyr (bottom). The colors represent the number density of the gas, while the white lines indicate the iso-contours of the temperature at T= 50, 75, 100, and 300 K.}
    \label{fig:density_temp_time}
\end{figure}

\begin{figure}
    \centering
    \includegraphics[width=0.5\textwidth, trim=0cm 0cm 0cm 0cm]{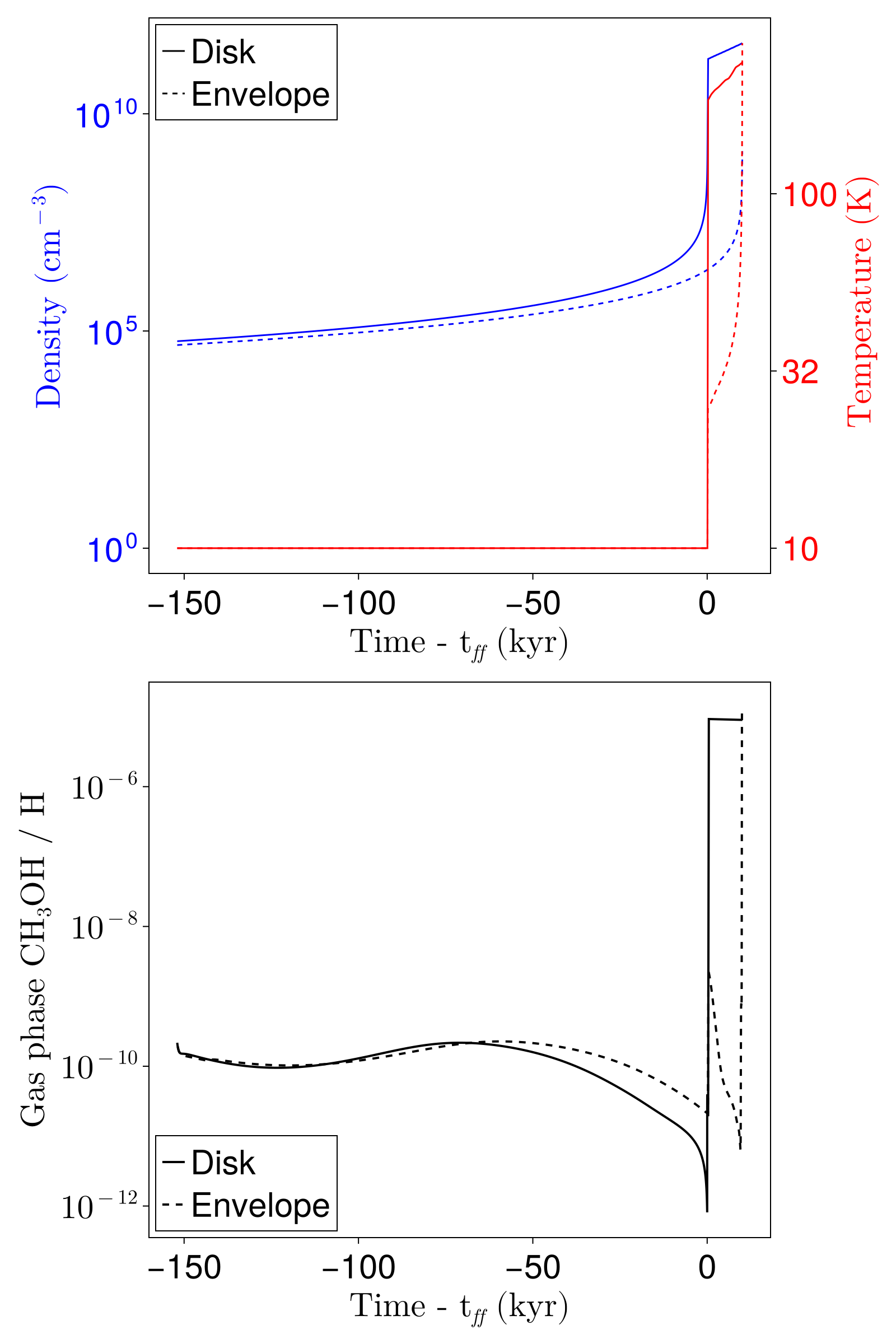}
    \includegraphics[width=0.48\textwidth, trim=2cm 0cm 0cm 0cm,clip]{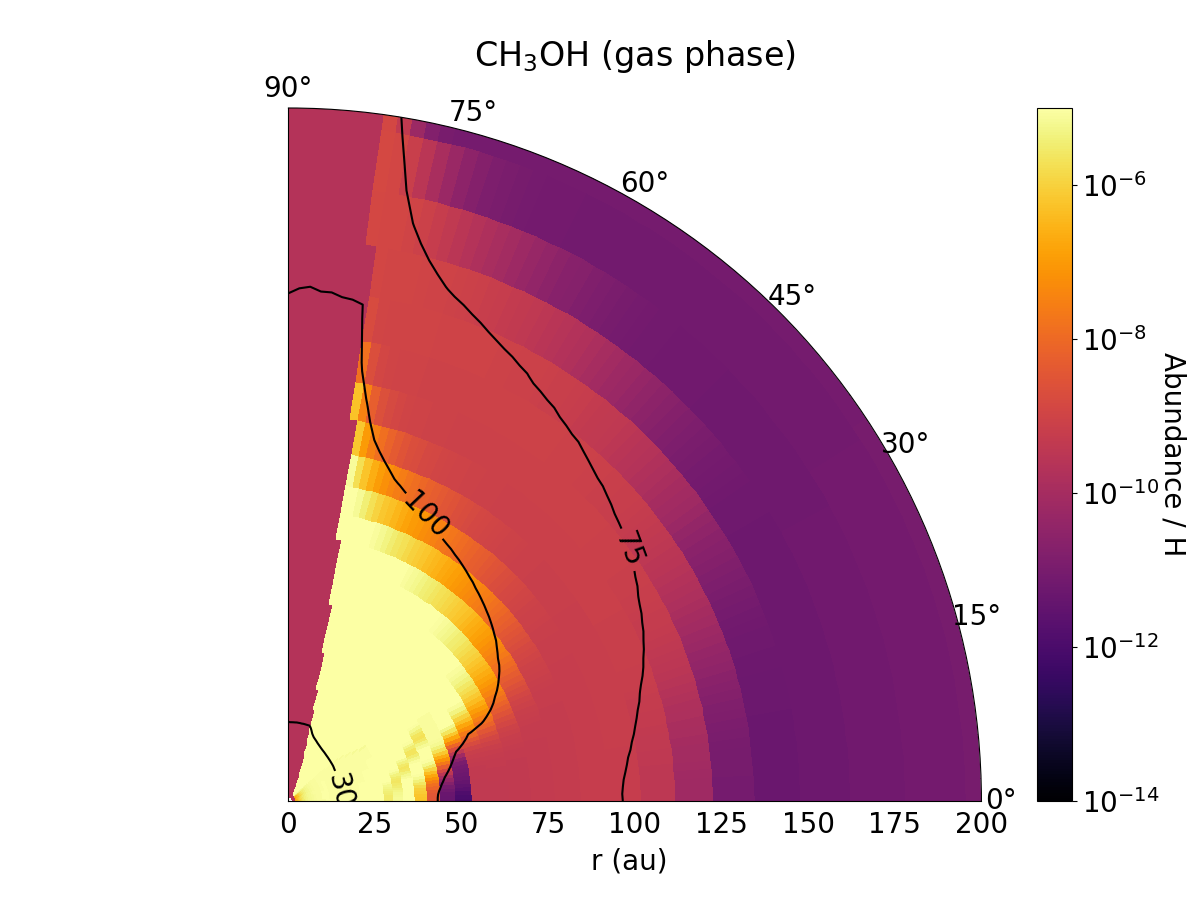}
    \caption{Upper panel: Density (blue) and temperature (red) history as a function of time for two particles. The solid and dashed lines represent particles located in the inner disk ($x,y = 19.8,1.4$ au), and in the inner envelope ($x,y = 14.0,14.0$ au), respectively, at their last time step at $t=\tff + 10$~kyr. Middle panel: Time evolution of the gas phase abundance of CH$_3$OH for the two particles. Bottom panel: Reconstructed map (magnified) of relative gas phase abundances of CH$_3$OH, 10 kyr after the formation of the central object. The black contours represent temperatures of 50, 75, 100, and 300 K.}
    \label{fig:part_traj_10}
\end{figure}

\begin{figure}
    \centering
    \includegraphics[width=0.5\textwidth, trim=0cm 0cm 0cm 0cm]{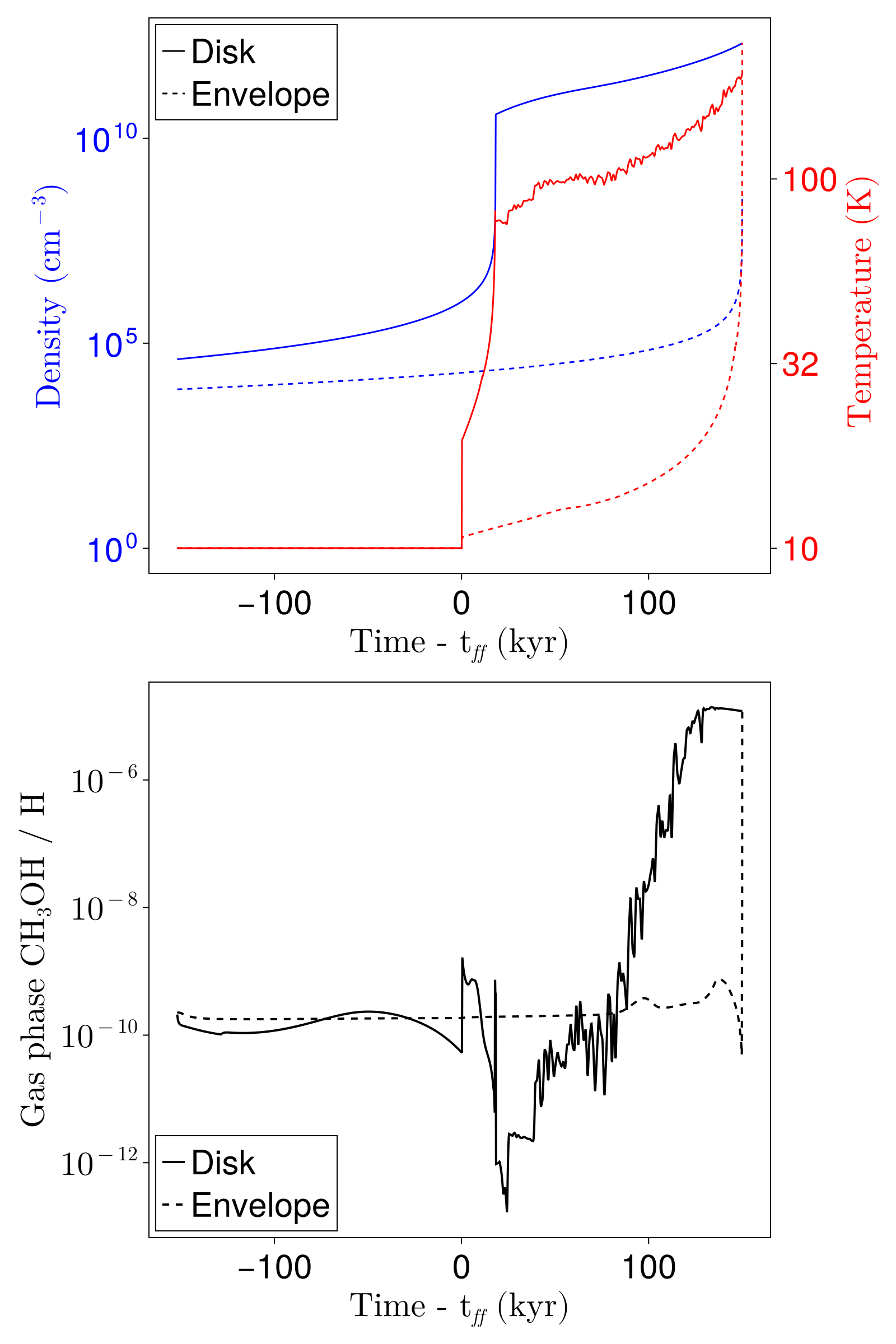}
    \includegraphics[width=0.48\textwidth, trim=2cm 0cm 0cm 0cm,clip]{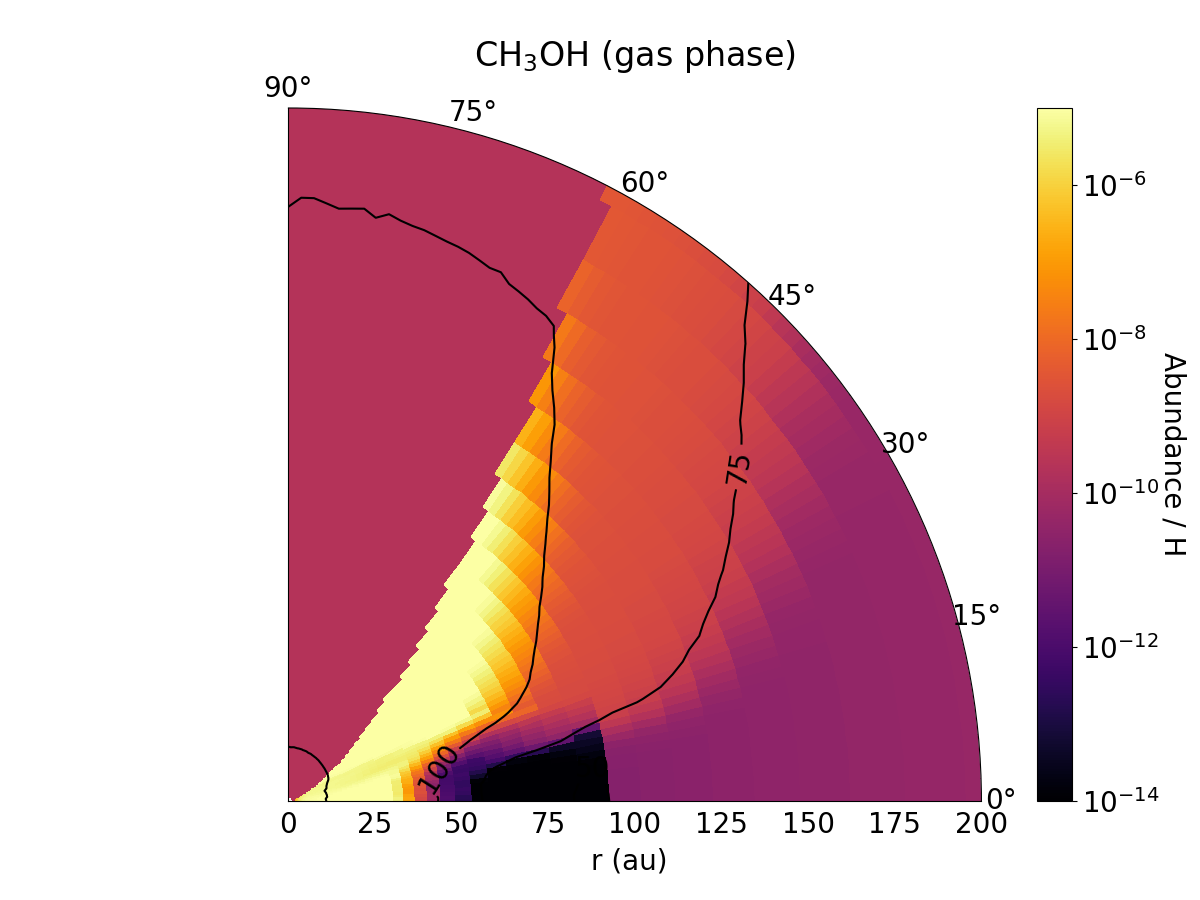}
    \caption{Same as Fig. \ref{fig:part_traj_10}, but at $t=\tff + 150$~kyr.}
    \label{fig:part_traj_150}
\end{figure}

Figure \ref{fig:density_temp_time} represents density and temperature maps of the simulation at $t=\tff+10$ and $t=\tff+150$ kyr. As the envelope mass is being accreted in the central region, the disk expands and the outflow widens. The increasing luminosity of the central object pushes away the temperature isocontours in the envelope. However, the protostellar radiation hardly penetrates farther into the disk, whose outer regions are shielded by the high disk density. The upper panel of Fig. \ref{fig:part_traj_10} displays the density and temperature history of two particles, one located in the envelope and one in the disk at $t=\tff+10$~kyr, both at a distance of $\sim 20$ au from the central object. The density of both particles starts at $\sim 10^5$ cm$^{-3}$ and slowly increases to $\sim 10^{7}$ cm$^{-3}$, until they reach the inner envelope ($\lesssim 100$ au). One of the particles (in solid lines) then enters the disk and experiences a density jump, before slowly drifting inward until the final time, 10 kyr later. The other particle sees its density increase faster as it draws close to the central object in the final kyr. The temperature of both particles is initially constant at $T=T_\mathrm{mc}=10$ K. The particles are then close enough to the central object at its formation to experience a jump of temperature, at $t-\tff=0$~kyr. The disk particle immediately reaches $T \sim 120$ K, and slowly heats up as it travels inward in the disk. The envelope particle is farther away at the formation of the central object, and its temperature only jumps to $T\sim 30$ K, and then heats up increasingly fast the following 10 kyr until it reaches $T \sim 120$ K.
Figure \ref{fig:part_traj_150} is the same as Fig. \ref{fig:part_traj_10} for $t=\tff+150$~kyr. The major difference is that those two particles are farther away from the central object at its formation, and experience smaller temperature jumps ($T \lesssim 20$ K). The disk particle reaches $T \approx 80$ K at its entry in the disk, and slowly heats up to $T \approx 120$ K over the following 150 kyr.

In APE, the formation of the central object is instantaneous at $t=\tff$. MHD simulations, on the other hand, show that a first hydrostatic core is first formed \citep{Larson1969}, and collapses into a protostar after less than a thousand years \citep[e.g.,][]{Bhandare2018}. This step is much shorter than the free-fall timescale, which is why we neglected it in our study.

\begin{table}
  \caption{Disk and central object properties at different times after the formation of the central object.}
  \label{tab:disk_star_prop}
\centering
\begin{tabular}{lllll}
\hline\hline
  Age (kyr) & 10 & 50 & 100 & 150\\
\hline
 $M_\mathrm{env}$ (M$_\odot$) & 1.64 & 1.12 & 0.54 & 0.05\\
 $\rd$ (au) & 52.6 & 71.0 & 83.9 & 92.3\\
 $M_\mathrm{disk}$ (M$_\odot$) & 0.082 & 0.20 & 0.34 & 0.45\\
 $M_*$ (M$_\odot$) & 0.28 & 0.68 & 1.12 & 1.50\\
 $R_*$ (R$_\odot$) & 2.96 & 3.6 & 5.25 & 5.35\\
 $L_*$ (L$_\odot$) & 29.8 & 57.9 & 61.7 & 72.8\\
 $T_*$ (K) & 7848 & 8387 & 7063 & 7290\\
 \hline
\end{tabular} 
\end{table}

\subsection{The example of the evolution of methanol} \label{sec:methanol}

\begin{figure*}
    \centering
    \sidecaption
    \includegraphics[width=12cm, trim=0cm 0cm 0cm 0cm]{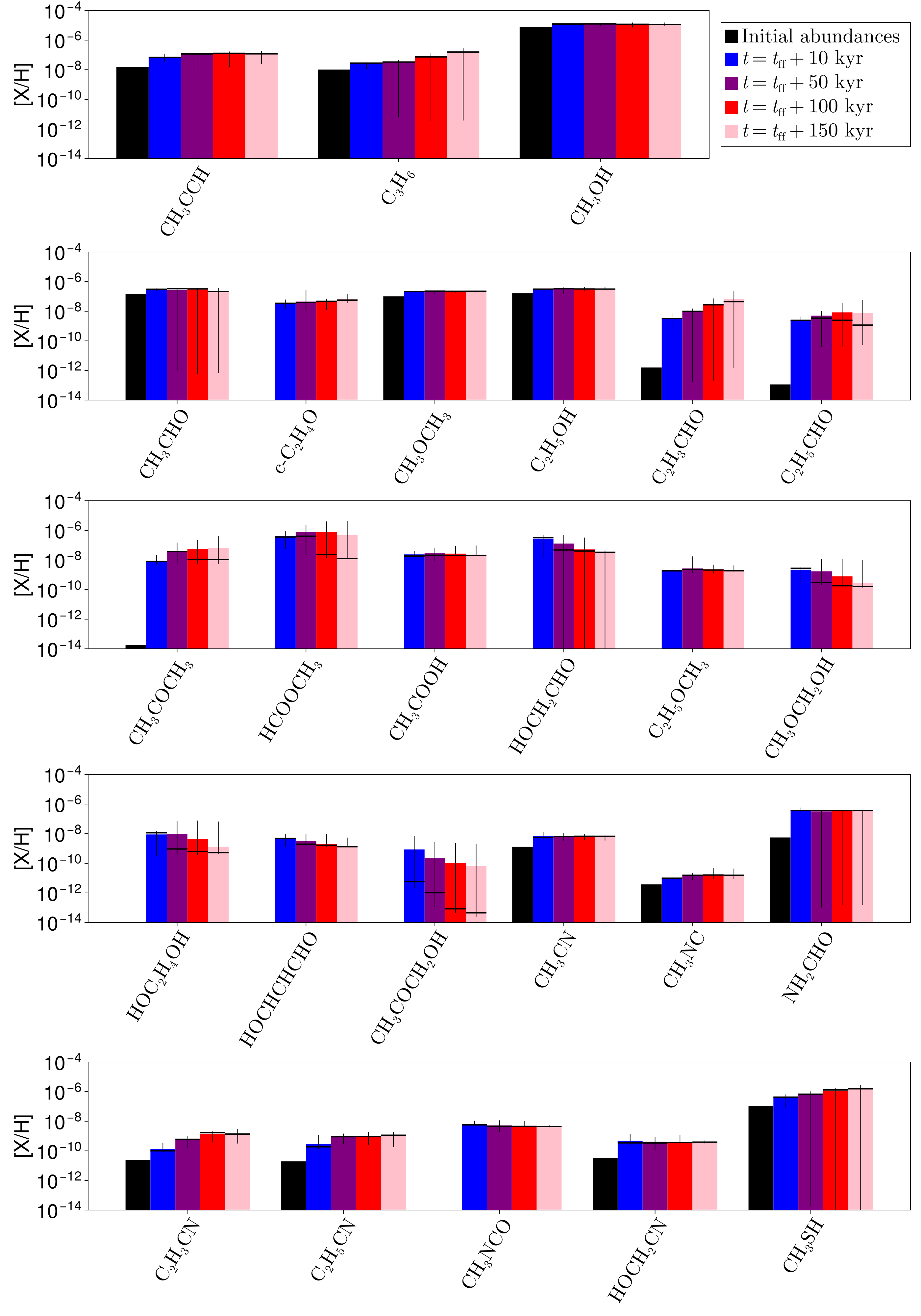}
    \caption{Relative abundance of several species in the midplane of the disk where $T<300$ K, at $t=\tff +10, 50, 100$, and $150$~kyr (colored bars). The height of the colored bar indicates the volume-averaged abundance over the midplane of the disk. The vertical black lines extends from the minimum to the maximum value. The horizontal black lines represent a median value above (and below) which 50\% of the volume of the gas stands. The black bars represent the initial abundances, and are not displayed if the initial abundances are lower than $10^{-14}$.}
    \label{fig:barplot_setup_abundances}
\end{figure*}

The density and temperature evolution are then used as input for the chemical calculations. As illustration, the central panels of Figs. \ref{fig:part_traj_10} and \ref{fig:part_traj_150} show the time evolution of the abundance (relative to H) of methanol (CH$_3$OH) in the gas phase for the same particles as in the upper panel. Methanol is present mainly in grains at low temperatures and sublimates at $T \gtrsim 100$ K. The abundances vary negligibly in the isothermal low-density phase, and start to decrease as the density increases around $t-\tff=-50$~kyr. At the formation of the central object, the jump in temperature causes a partial sublimation of the molecules frozen on the grains for the envelope particle in the $t=\tff+10$ kyr case (Fig. \ref{fig:part_traj_10}), and for the disk particle in the $t=\tff+150$ kyr case (Fig. \ref{fig:part_traj_150}). The sublimation is total for the disk particle in the $t=\tff+10$ kyr case as the temperature immediately exceeds 100 K, and the methanol remains fully sublimated in the gas phase until the end of the simulation. The disk particle in the $t=\tff+150$ kyr case experiences a density jump at the disk entry, which causes a sudden decrease of methanol gas phase abundance. As the particle drifts inward in the disk, the increasing temperature provokes a slow sublimation of CH$_3$OH in the gas phase over time. The sublimation is fast for both envelope particles, which reach high-temperature regions at the very end of their run. The final abundances in the disk and in the envelope are similar, as all CH$_3$OH molecules present on grains have sublimated.

We perform similar calculations for all particles in the simulation box, and use the final abundances of each particle to reconstruct abundance maps. We display the resulting CH$_3$OH maps at $t=\tff+10$~kyr and $t=\tff+150$~kyr in the bottom panels of Figs. \ref{fig:part_traj_10} and \ref{fig:part_traj_150}. There is a clear correlation between the regions of high methanol abundance and the temperature isocontours. Those maps only display the gas phase abundance of methanol; the total gas+ice abundance is roughly constant throughout the collapse (see next paragraphs and Fig. \ref{fig:barplot_setup_abundances}); therefore, the two maps effectively illustrate the thermal sublimation of the species.

\subsection{Evolution of the composition in COMs in the disk midplane} \label{sec:time-COM}

In the rest of this section, we analyze how COMs evolve through time from the prestellar phase to the Class I stage. We considered only cells belonging to the midplane of the disk as it is the location of planet formation. We excluded the cells with a temperature above 300~K ($r \lesssim 15$ au). In addition, COMs in the innermost region (a few au) may be subjected to UV dissociation, a process that was not included in our model. We focused on the evolution of COMs that have been detected in protostellar sources \citep[e.g.,][]{Cazaux2003,Lykke2017,Manigand2021,Coutens2022}. We excluded only HC$_5$N whose grain chemistry is not precise in our network. The 26 species we considered are listed in Table \ref{tab:COMs}.

\begin{table}
  \caption{COMs considered in this study, with their binding energy $E_b$. In each category, the species are ranked by atomic mass.}
  \label{tab:COMs}
\centering
\begin{tabular}{llc}
\hline\hline
  Formula & Name & $E_b/\kb$ (K)\\
\hline
\multicolumn{3}{c}{Hydrocarbons}\\
\hline 
 CH$_3$CCH & Propyne & 4000\\
 C$_3$H$_6$ & Propylene & 3600\\
\hline 
\multicolumn{3}{c}{O-bearing COMs}\\
\hline 
 CH$_3$OH & Methanol & 5000\\
 CH$_3$CHO & Acetaldehyde & 5500\\
 c-C$_2$H$_4$O & Ethylene Oxide & 5000\\
 CH$_3$OCH$_3$ & Dimethyl ether & 4100\\
 C$_2$H$_5$OH & Ethanol & 5500\\
 C$_2$H$_3$CHO & Propanal & 5500\\
 C$_2$H$_5$CHO & Propionaldehyde & 4500\\
 CH$_3$COCH$_3$ & Acetone & 4100\\
 HCOOCH$_3$ & Methyl formate & 3500\\
 CH$_3$COOH & Acetic acid & 6000\\
 HOCH$_2$CHO & Glycolaldehyde & 5000$^a$\\
 C$_2$H$_5$OCH$_3$ & Ethyl methyl ether & 5000\\
 CH$_3$OCH$_2$OH & Methoxymethanol & 6000\\
 HOC$_2$H$_4$OH & Ethylene glycol & 6000\\
 HOCHCHCHO & 3-hydroxypropenal & 6000\\
 CH$_3$COCH$_2$OH & Hydroxyacetone & 6000\\
\hline 
\multicolumn{3}{c}{N-bearing COMs}\\
\hline 
 CH$_3$CN & Methyl cyanide & 4100\\
 CH$_3$NC & Methyl isocyanide & 4000$^a$\\
 NH$_2$CHO & Formamide & 6500\\
 C$_2$H$_3$CN & Vinyl Cyanide & 3000\\
 C$_2$H$_5$CN & Ethyl cyanide & 4400\\
 CH$_3$NCO & Methyl isocyanate & 4700\\
 HOCH$_2$CN & Glycolonitrile & 4000$^a$\\
\hline 
\multicolumn{3}{c}{S-bearing COMs}\\
\hline 
 CH$_3$SH & Methanethiol & 4100\\
\hline 
\end{tabular} 
\tablefoot{$^a$ Those values are estimations based on species with similar behaviors.}
\end{table}

Figure \ref{fig:barplot_setup_abundances} displays the average, minimum, maximum and median abundance of those species in the midplane of the disk, for protostellar ages of $10$, $50$, $100$, and $150$ kyr. The median has been defined so that, in half of the total volume of the gas, the species abundance (/H) is lower than this value. With this definition, particles at large radii represent a larger fraction of the volume of the disk, and therefore weigh more significantly in the calculation, than particles at small radii. The plotted abundances are the total abundances (gas phase + grain surface + grain mantle), so that variations in abundances represent the actual formation and destruction of the species (not a freeze-out on grains or a sublimation). The full profiles of abundance of the disk midplane are displayed in Appendix \ref{ann:profiles} (see Fig. \ref{fig:grid_times}).

\subsubsection{Inheritance and formation pathways}

For 16 of the 26 molecules, the average disk abundances at $t=\tff+10~$kyr are more than one order of magnitude higher than their initial abundance, meaning that they are primarily formed during the collapse: c-C$_2$H$_4$O, C$_2$H$_3$CHO, C$_2$H$_5$CHO, CH$_3$COCH$_3$, HCOOCH$_3$, CH$_3$COOH, HOCH$_2$CHO, C$_2$H$_5$OCH$_3$, CH$_3$OCH$_2$OH, HOC$_2$H$_4$OH, HOCHCHCHO, CH$_3$COCH$_2$OH, NH$_2$CHO, C$_2$H$_5$CN, CH$_3$NCO, and HOCH$_2$CN. This list includes the 11 heaviest O-bearing COMs and the 3 heaviest N-bearing COMs of the selected species.

\begin{figure}
    \centering
    \includegraphics[width=0.49\textwidth, trim=0cm 0cm 0cm 0cm]{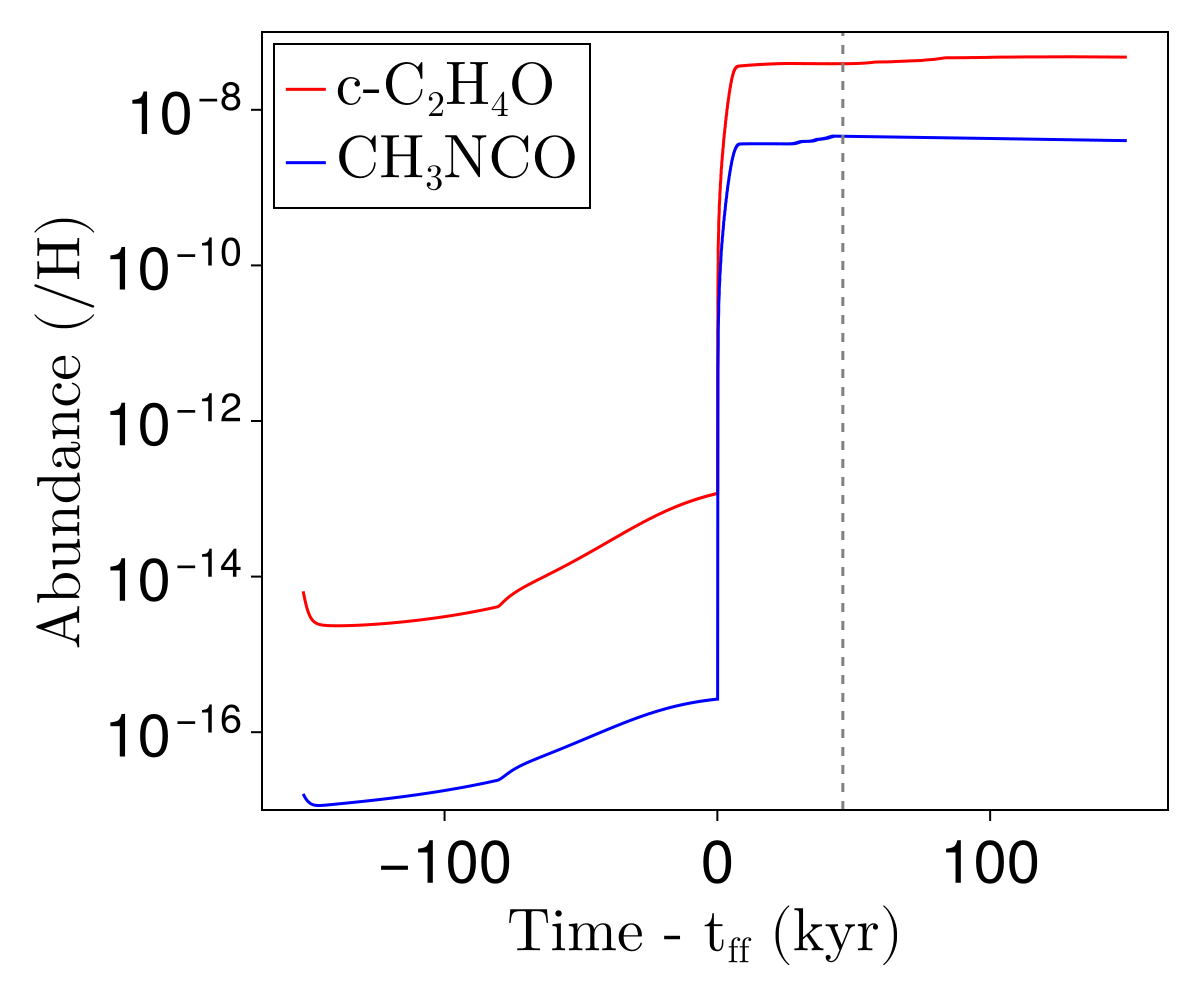}
    \caption{Time evolution of the total abundance (gas + grain surface + grain mantle) of c-C$_2$H$_4$O (red) and CH$_3$NCO (blue) for a particle ending in the midplane of the disk at a distance of 50 au from the central object at $t=\tff+150$~kyr. The central object forms at $t-\tff=0$~kyr, and the dashed vertical grey line represents the entry of the particle in the disk.}
    \label{fig:abund_evol_com}
\end{figure}

Figure \ref{fig:abund_evol_com} displays the evolution of the total abundance of c-C$_2$H$_4$O and CH$_3$NCO for the particle ending in the midplane at $r=50$~au at $t=\tff+150$~kyr. Initially, the abundances slowly increase before the central object formation due to the rise in density. The abundances then jump by several orders of magnitude at the formation of the central object, due to the rise in temperature in the system, before the disk entry. All COMs primarily formed during the collapse follow a similar scenario.

Two molecules are formed in majority during the prestellar phase (i.e., their initial abundance is at least 50\% of the average abundance at $t=\tff+10$~kyr): CH$_3$OH and C$_2$H$_5$OH. 
The eight remaining species are mainly formed during the collapse but are still significantly formed during the prestellar phase (i.e., their initial abundance is within an order of magnitude from the average disk abundance at $t=\tff+10$~kyr): CH$_3$CCH, C$_3$H$_6$, CH$_3$CHO, CH$_3$OCH$_3$, CH$_3$CN, CH$_3$NC, C$_2$H$_3$CN, and CH$_3$SH.

Most of the COMs formed during the collapse are produced from radicals reacting on grains, namely c-C$_2$H$_4$O, CH$_3$COOH, HOCH$_2$CHO, C$_2$H$_5$OCH$_3$, CH$_3$OCH$_2$OH, HOC$_2$H$_4$OH, HOCHCHCHO, CH$_3$COCH$_2$OH, NH$_2$CHO, C$_2$H$_5$CN, CH$_3$NCO, and HOCH$_2$CN. These formation pathways require radicals to be mobile on the grains, which in turn necessitates higher temperatures than those found during the prestellar phase.

The inherited COMs CH$_3$OH and C$_2$H$_5$OH, as well as CH$_3$CHO and CH$_3$OCH$_3$, are also produced on grains. They primarily or partially form by successive hydrogenations of species derived from CO. This occurs at low temperature because CO needs to be frozen on grains, which results in a significant formation of those species at early stages. However, the reservoir accumulated during the prestellar phase is not necessarily preserved throughout the collapse. For example, CH$_3$OH is continuously destroyed in grain mantles either by reaction with H or by photodissociation due to UV photons induced by cosmic rays, thus forming H$_2$CO+H+H or the radicals CH$_3$+OH. It is however rapidly re-formed and its total abundance does not vary significantly throughout the collapse. The total amount of CH$_3$OH destroyed and reformed approximates the value of its final abundance. C$_2$H$_5$OH exhibits a similar behavior, but only half of its final abundance has been destroyed and reformed throughout the collapse, meaning that at least a part of its initial reservoir remained intact during the journey.

CH$_3$CN and CH$_3$NC are also both formed in the grains, from two successive hydrogrenations of HCCN. HCCN is produced both in the gas phase or by N+C$_2$H in ices.
Similarly, CH$_3$SH is formed by the attachment of S onto hydrocarbon radicals in the gas phase, then from hydrogenations in ices.
The two hydrocarbons CH$_3$CCH and C$_3$H$_6$ form both in the gas phase, mostly from C$_3$H$_5^+$ and C$_3$H$_7^+$, as well as on the grains from other hydrocarbons. 

C$_2$H$_3$CHO, C$_2$H$_5$CHO, CH$_3$COCH$_3$, and HCOOCH$_3$ are produced from the recombination of ions in the gas phase. C$_2$H$_3$CHO is formed from C$_2$H$_3$CHOH$^+$, itself forming from gas phase CO. C$_2$H$_5$CHO, and CH$_3$COCH$_3$ are formed from C$_2$H$_5$CHOH$^+$ or CH$_3$COHCH$_3^+$, both forming from C$_2$H$_4$ in the gas phase. HCOOCH$_3$ is produced by the recombination of HCOHOCH$_3^+$, coming from H$_2$CO. The production of those four molecules can therefore only occur above the sublimation temperature of CO, C$_2$H$_4$, or H$_2$CO (binding energies of 1300~K, 2500~K, and 3200~K in the network, respectively).
C$_2$H$_3$CN is also majorly produced in the gas phase, by a reaction between CN and C$_2$H$_4$.

\subsubsection{Time evolution}

Figure \ref{fig:barplot_setup_abundances} also displays the abundance evolutions from $t=\tff+10$~kyr to $t=\tff+150$~kyr, which are summarized in Table \ref{tab:spec_evol} for each COM. 
15 species experience small variations (less than a factor of 2) of their average and median abundance, namely CH$_3$CCH, CH$_3$OH, CH$_3$CHO, c-C$_2$H$_4$O, CH$_3$OCH$_3$, C$_2$H$_5$OH, HCOOCH$_3$, CH$_3$COOH, HOCH$_2$CHO, C$_2$H$_5$OCH$_3$, CH$_3$CN, CH$_3$NC, NH$_2$CHO, CH$_3$NCO, and HOCH$_2$CN. The constant abundance of methanol is in accordance with previous studies \citep{Drozdovskaya2016,Coutens2020}. Among those species, CH$_3$OH, CH$_3$OCH$_3$, C$_2$H$_5$OH, and CH$_3$CN are uniformly distributed throughout the disk, with less than a factor of 3 between their maximum and minimum value. The small spread of those species is consistent with the results of \citet{Drozdovskaya2016}, who investigated the evolution of chemical species in a protoplanetary disk until the Class II phase. On the contrary, the abundances of CH$_3$CHO, HOCH$_2$CHO, and NH$_2$CHO span a large range of values in the disk. Their abundance decreases by several orders of magnitude in the inner 30 au after $t=\tff+50$ or $100$~kyr (see fig. \ref{fig:grid_times}), which also happens to C$_3$H$_6$, C$_2$H$_3$CHO, C$_2$H$_5$CHO, and CH$_3$SH. This behavior is also observed in \citet{Drozdovskaya2016} for CH$_3$CHO. Among those species, CH$_3$CHO, C$_2$H$_3$CHO, C$_2$H$_5$CHO, and HOCH$_2$CHO are destroyed on grains by exchanging a H atom with CH$_3$O, which produces CH$_3$OH. C$_3$H$_6$ and NH$_2$CHO are destroyed on grains by reactions with oxygen atoms. Those behaviors from CH$_3$O and O, possibly an excess of formation or a lack of destruction, are difficult to explain in details, because their formation and destruction are involved in many inter-dependent reactions.

Seven species become more abundant in average between $t=\tff+10$~kyr and $t=\tff+150$~kyr: C$_3$H$_6$, C$_2$H$_3$CHO, C$_2$H$_5$CHO, CH$_3$COCH$_3$, C$_2$H$_3$CN, C$_2$H$_5$CN, and CH$_3$SH. Five other species, only heavy O-bearing COMs, become less abundant: HOCH$_2$CHO, CH$_3$OCH$_2$OH, HOC$_2$H$_4$OH, HOCHCHCHO, and CH$_3$COCH$_2$OH. The variations are monotonous in time. The amplitude of the variations between the $t=\tff+10$~kyr and $t=\tff+150$~kyr ranges from a factor of a few (e.g., C$_2$H$_5$CN) to more than one order of magnitude (CH$_3$COCH$_2$OH). The maximum and minimum abundance of those increasing and decreasing species mostly follow similar trends, except for the molecules experiencing a depletion in the inner 30~au (C$_3$H$_6$, CH$_3$CHO, C$_2$H$_3$CHO, C$_2$H$_5$CHO, HOCH$_2$CHO, NH$_2$CHO, and CH$_3$SH).

The median abundances generally follow the variations of the average abundances, with one exception: HCOOCH$_3$. Although its overall abundance varies by less than a factor of 2 between $t=\tff+10$~kyr and $t=\tff+150$~kyr, its median abundance decreases by nearly two orders of magnitude. This is due to the radial expansion of the disk and the low concentration of this species in the outer regions, which makes up the majority of the volume of the disk.

\begin{table}
  \caption{Time-evolution of the average abundances of COMs in the disk between $t=\tff+10$~kyr and $t=\tff+150$~kyr.}
  \label{tab:spec_evol}
\centering
\begin{tabular}{lll}
\hline\hline
Increasing & Constant & Decreasing \\
\hline
C$_3$H$_6$  & CH$_3$CCH  &  {\color{red} HOCH$_2$CHO}\\
{\color{red} C$_2$H$_3$CHO}  & {\color{red} CH$_3$OH}  & {\color{red} CH$_3$OCH$_2$OH} \\
{\color{red} C$_2$H$_5$CHO}  & {\color{red} CH$_3$CHO}  &  {\color{red} HOC$_2$H$_4$OH}\\
{\color{red} CH$_3$COCH$_3$} & {\color{red} c-C$_2$H$_4$O}  &  {\color{red} HOCHCHCHO}\\
{\color{blue} C$_2$H$_3$CN} & {\color{red} CH$_3$OCH$_3$}  & {\color{red} CH$_3$COCH$_2$OH} \\
{\color{blue} C$_2$H$_5$CN}  & {\color{red} C$_2$H$_5$OH} &  \\
{\color{brown} CH$_3$SH}  & {\color{red} HCOOCH$_3$}  &  \\
  & {\color{red} CH$_3$COOH}  &  \\
  & {\color{red} HOCH$_2$CHO}  &  \\
  & {\color{red} C$_2$H$_5$OCH$_3$}  &  \\
  & {\color{blue} CH$_3$CN}  &  \\
  & {\color{blue} CH$_3$NC}  &  \\
  & {\color{blue} NH$_2$CHO}  &  \\
  & {\color{blue} CH$_3$NCO}  &  \\
  & {\color{blue} HOCH$_2$CN}  &  \\
\hline
\end{tabular} 
\tablefoot{Species in the ``Constant'' column experience abundance variations of less than a factor of 2 at all considered times. Species are color-coded according to their nature: hydrocarbons in black, O-bearing species in red, N-bearing species in blue and the S-bearing species in brown.}
\end{table}

\subsubsection{Dominant phases}

\begin{figure}
    \centering
    \includegraphics[width=0.49\textwidth, trim=0cm 0cm 0cm 0cm]{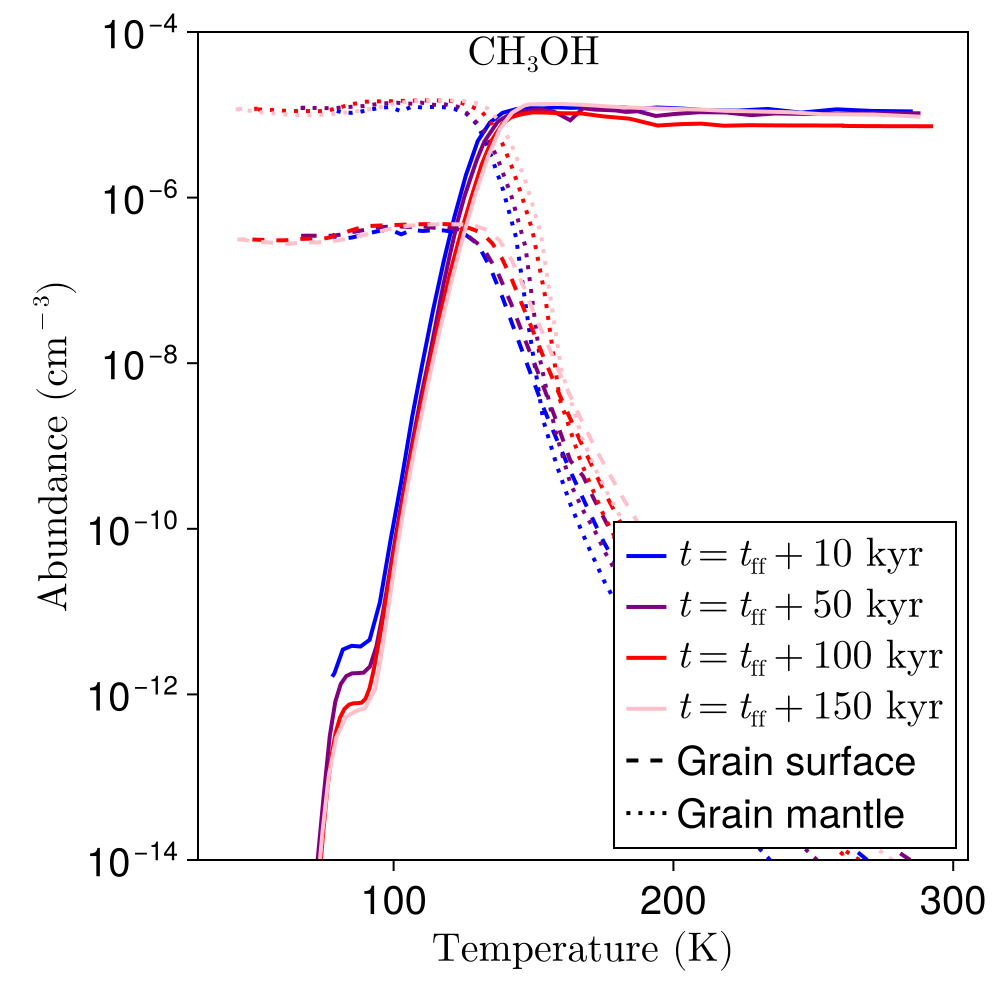}
    \caption{Abundance of CH$_3$OH as a function of temperature in the midplane of the disk. The different colors indicate different protostellar ages, as in Fig. \ref{fig:barplot_setup_abundances}. The solid lines represent the gas phase, the dashed lines represent the grain surface phase, and the dotted lines represent the grain mantle phase.}
    \label{fig:phases}
\end{figure}

The reservoir of COMs in the disk is distributed across the three phases considered by Nautilus: the gas, the mantle of grains, and the surface of grains. The phase that dominates the distribution of a given species is uniquely dependent on the temperature.
Figure \ref{fig:phases} displays the abundance of the three phases of CH$_3$OH as a function of temperature in the midplane of the disk. At low temperature ($\lesssim 100$~K), the mantle phase dominates in abundance, followed by the grain surface 1 to 2 orders of magnitude below, while the presence of CH$_3$OH in the gas phase is negligible. This distribution changes around 100~K, temperature at which the COMs sublimate. Above 120~K, nearly all the CH$_3$OH is in the gas phase, while the abundances in the grain mantle and on the surfaces decrease gradually with temperature at similar values above $\sim 150$~K.

All species behave similarly to CH$_3$OH. The gas and ice (mantle+surface) phases abundance profiles are plotted in Figs. \ref{fig:grid_times_phases} and \ref{fig:grid_preabund_phases}. The ice phase significantly prevails over the gas phase at temperatures below the sublimation point. Above it, the gas phase becomes dominant. Table \ref{tab:COMs} lists the binding energies of all species in our chemical network.

\subsection{Sensitivity to the physical conditions}

\begin{figure*}
    \centering
    \sidecaption
    \includegraphics[width=12cm, trim=0cm 0cm 0cm 0cm]{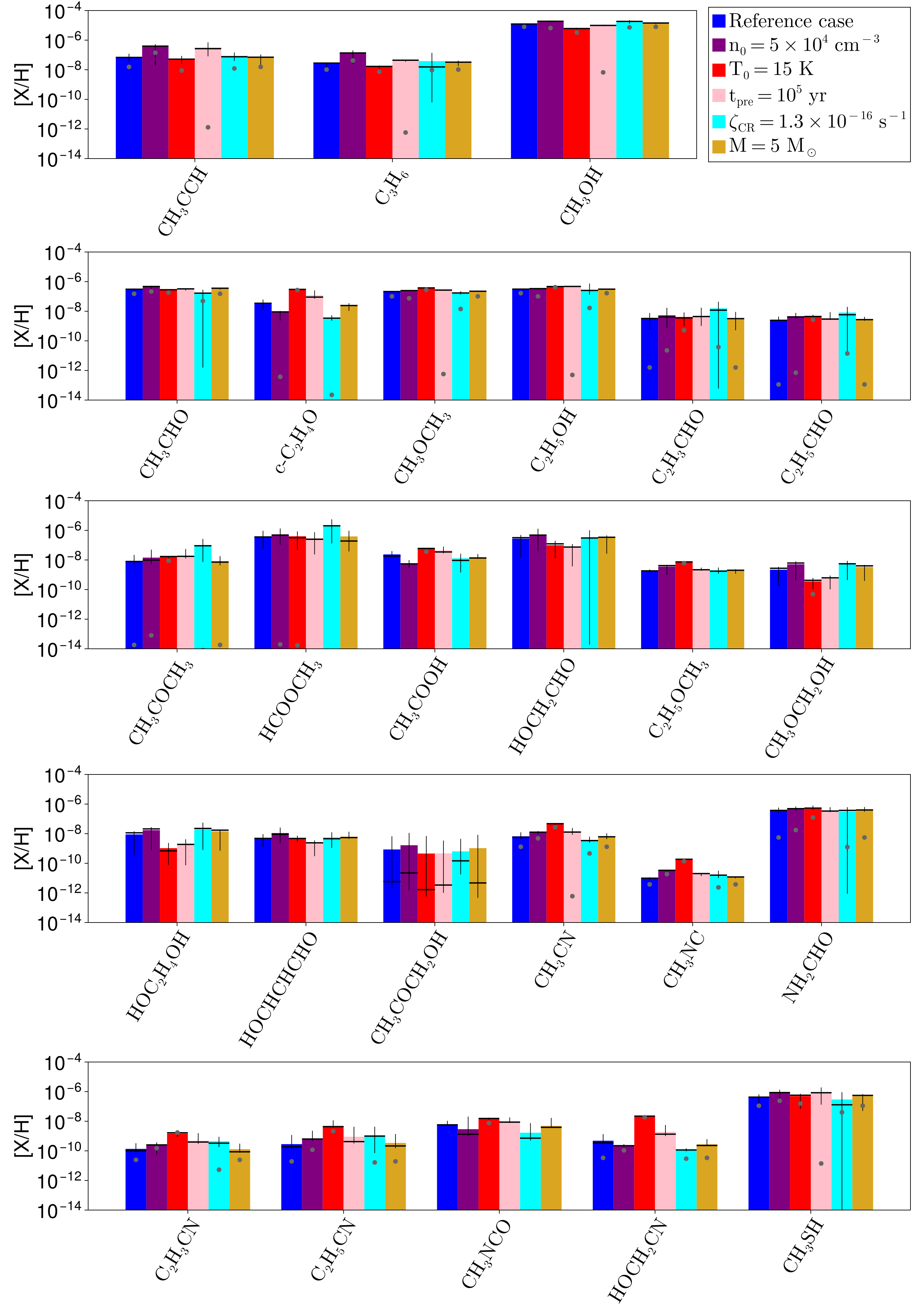}
    \caption{Same as Fig. \ref{fig:barplot_setup_abundances}, but for different physical conditions at $t=\tff+10$~kyr. The gray points represent the initial abundances, which are not displayed for values lower than $10^{-14}$.}
    \label{fig:barplot_abundini_abundances}
\end{figure*}

As detailed in the previous sections, COMs can be formed in the prestellar phase or during the collapse. Changes of physical conditions in the prestellar phase may therefore alter the abundances used as initial conditions for the simulations, which can be reflected on the final abundances despite identical final physical conditions \citep{Coutens2020}. Similarly, different physical conditions during the collapse impact the chemical evolution of the infalling gas. In this section, we analyze how the disk abundances are affected by the variations of different physical parameters of the cloud.

Changing only one parameter at a time, we decreased the prestellar phase duration from $t_\mathrm{pre}=10^6$~yr to $t_\mathrm{pre}=10^5$~yr, increased the prestellar core density from $n_0=10^4$~cm$^{-3}$ to $n_0=5\times 10^4$~cm$^{-3}$, increased the cosmic-ray ionization rate in the prestellar core and throughout the collapse from $\zeta_\mathrm{CR}=1.3 \times 10^{-17}$~s$^{-1}$ to $\zeta_\mathrm{CR}=1.3 \times 10^{-16}$~s$^{-1}$, and increased the temperature of the molecular cloud from $T_0=T_\mathrm{mc}=10$~K to $T_0=T_\mathrm{mc}=15$~K. The changes in duration $t_\mathrm{pre}$ and density $n_0$ only apply to the prestellar phase, while $T_\mathrm{mc}$ and $\zeta_\mathrm{CR}$ are changed for both the prestellar phase and the collapse. We additionally performed a simulation with a $M=5$~M$_\odot$ cloud. In that case, the envelope density is lower because the initial Bonnor-Ebert sphere is more extended, and the free-fall time is 380~kyr. The particles therefore spend more than twice the time in the envelope before reaching the central region, compared to the $M=2$~M$_\odot$ case. For all five parameter variations, we focused on a time $10$~kyr after the formation of the central object. The results are presented in Fig. \ref{fig:barplot_abundini_abundances} in the same manner as in Fig. \ref{fig:barplot_setup_abundances}, and the abundance profiles are displayed in Fig. \ref{fig:grid_preabund}.

\subsubsection{Sensitivity to the prestellar phase density}

Increasing the density of the prestellar phase promotes the formation of the majority of COMs. Most species experience an increase in initial abundance by a factor of a few, namely CH$_3$CCH, C$_3$H$_6$, CH$_3$CHO, C$_2$H$_5$CHO, CH$_3$COCH$_3$, CH$_3$CN, NH$_2$CHO, C$_2$H$_3$CN, C$_2$H$_5$CN, HOCH$_2$CN, and CH$_3$SH. This category includes both simpler and more complex COMs. Only three molecules see their initial abundance increased by more than one order of magnitude (up to a factor of 54): c-C$_2$H$_4$O, C$_2$H$_3$CHO, and HCOOCH$_3$. The molecules CH$_3$OH, CH$_3$OCH$_3$, and C$_2$H$_5$OH, have lower initial abundances by 20\% to 40\%. The other molecules have initial abundances lower than $10^{-14}$. 

Although the abundances at $t=\tff+10$~kyr are impacted by this change in initial condition, the differences are attenuated in the disk. Nearly all (22) COMs have final average and median abundances that are a factor of 1.1 to 5 higher than the reference case. 
Only four species have reduced final abundances, by a factor of 2 to 5 on the average and the median; namely c-C$_2$H$_4$O, CH$_3$COOH, CH$_3$NCO, and HOCH$_2$CN. 
The median abundances vary by the same factor (within 50\%) as the average abundances, suggesting a somewhat uniform changes throughout the disk. The only exceptions are CH$_3$COCH$_2$OH and CH$_3$NCO, for which the average abundance vary by a factor of $\sim 2$ while the median abundance vary by a factor of $\sim 4$.

\subsubsection{Sensitivity to the temperature of the molecular cloud}

Changing the molecular cloud temperature alters the equilibrium between the gravity and the thermal pressure of a Bonner-Ebert sphere. Increasing the temperature from $T_\mathrm{mc}=10$~K to $T_\mathrm{mc}=15$~K results in a 1.5 times smaller cloud and a $1.5^{\frac{3}{2}}\approx 1.8$ shorter free-fall time (i.e., $\sim 83$ kyr in this case). However, as demonstrated in Sect. \ref{sec:methanol} and \ref{sec:time-COM}, particularly Fig. \ref{fig:abund_evol_com}, the abundance of COMs are primarily affected by density and temperature variations, at long and short timescales. 

Like the increase in density, this increase in temperature promotes the formation of COMs in the prestellar phase. The exceptions are again CH$_3$OH, but also the two hydrocarbons CH$_3$CCH and C$_3$H$_6$. The initial abundances of those three species are lower by a factor of 1.5 to 2.
Several species only experience an increase in initial abundance lower than one order of magnitude, namely CH$_3$CHO, CH$_3$OCH$_3$, C$_2$H$_5$OH, and CH$_3$SH. Another set of species, mostly the lightest N-bearing COMs, see an increase of one to two orders of magnitude, namely HCOOCH$_3$, CH$_3$CN, CH$_3$NC, NH$_2$CHO and C$_2$H$_5$CN. 
The most notable impact on initial abundances, however, is seen for species that have very low ($\lesssim 10^{-14}$) initial abundances in the reference case that increase above $10^{-10}$ with a higher temperature. The species in this case are c-C$_2$H$_4$O, CH$_3$COCH$_3$, CH$_3$COOH, C$_2$H$_5$OCH$_3$, CH$_3$OCH$_2$OH, and CH$_3$NCO.

The higher temperature results in 16 species with average and median abundances in the disk higher than the reference case at $t=\tff+10$~kyr. The increase  is less than a factor of 3, except for c-C$_2$H$_4$O (factor of $\sim 8$), C$_2$H$_5$OCH$_3$ ($\sim 4$), CH$_3$CN ($\sim 8$), CH$_3$NC ($\sim 20$), C$_2$H$_3$CN ($\sim 13$), C$_2$H$_5$CN ($\sim 20$), and HOCH$_2$CN ($\sim 55$). In those cases, except C$_2$H$_5$CN, the initial abundance is higher than the maximum disk abundance of the reference case (see Fig \ref{fig:grid_preabund}). This results in uniform abundance profiles at the value of the initial abundance. Eight of the remaining species have slightly lower average and median abundances than the reference case by factors of $\lesssim 2$, namely CH$_3$CCH, C$_3$H$_6$, CH$_3$OH, CH$_3$CHO, C$_2$H$_3$CHO, HCOOCH$_3$, HOCH$_2$CHO, and HOCHCHCHO. Two additional species, CH$_3$OCH$_2$OH and HOC$_2$H$_4$OH, are more significantly depleted with abundances close to one order of magnitude lower than the reference case at all radii. Similarly to the change of prestellar phase density, all changes of abundance profiles are uniform throughout the disk (i.e., the average and median abundances change by similar factors).

\subsubsection{Sensitivity to the prestellar phase duration}

Reducing the prestellar phase duration by a factor of 10 reduces the initial abundances of all COMs to values below $10^{-14}$, as they have less time to form from atomic abundances. The disk average and median abundances are however close to the reference case within a factor of 2 for 14 species, and within a factor of 5 for the 12 other. Eighteen species are more abundant than in the reference case, while eight are less abundant, mostly heavy O-bearing COMs: CH$_3$OH, HCOOCH$_3$, HOCH$_2$CHO, CH$_3$OCH$_2$OH, HOC$_2$H$_4$OH, HOCHCHCHO, CH$_3$COCH$_2$OH, and NH$_2$CHO. For every species, the change in abundance profile is nearly uniform throughout the disk.

\subsubsection{Sensitivity to the cosmic-ray ionization rate}

Increasing the cosmic-ray ionization rate leads to a general decrease of COM abundances. First, during the prestellar phase, where all but four species are produced in lower quantities, up to a factor of 12. Among species with initial abundances larger than $10^{-14}$ in the reference case, only two have larger initial abundances with an increased ionization rate: C$_2$H$_3$CHO with a factor of 24 and C$_2$H$_5$CHO with a factor of 122.

In the disk the largest differences in average abundance are experienced by c-C$_2$H$_4$O, CH$_3$COCH$_3$, and HCOOCH$_3$, which are more abundant by one order of magnitude compared to the reference case. The other species have average abundances that are higher or lower than the reference case by  factors of $\lesssim 4$. This increase in cosmic-ray ionization rate however significantly impacts the abundance profile of several species: C$_3$H$_6$, CH$_3$CHO, C$_2$H$_3$CHO, C$_2$H$_5$CHO, HOCH$_2$CHO, NH$_2$CHO, and CH$_3$SH. Despite small changes in average value, their abundances drops by 2 to 8 orders of magnitude between $r\approx 25$~au and $r\approx 40$ au. Those species are the same that experience depletion in the inner disk after $t=\tff+50$~kyr in the fiducial case (see Sect. \ref{sec:time-COM} and Fig. \ref{fig:grid_times}).

\subsubsection{Sensitivity to the cloud mass}

Changing the mass of the cloud impacts its free-fall time and the physical history of particles, but we assume that the initial abundances are the same as the reference case.
The abundance profiles in the disk nearly overlap for almost all species. 
The average and median abundances obtained with the two cloud masses are very close, with differences always lower than a factor of 2. The spread in abundances for each species, represented by the vertical black lines in Fig. \ref{fig:barplot_abundini_abundances}, is also only lightly affected. While the abundance profile of most species remains relatively flat, a few exceptions, namely C$_2$H$_3$CHO, HCOOCH$_3$, CH$_3$COCH$_2$OH, and C$_2$H$_5$CN, exhibit an outward shift of their profile of approximately 5 au. This shift corresponds to the displacement of isothermal contours, which results from the higher temperature of the central object.
This similarity in abundances for all species supports the idea that the time spent in the envelope has only a little impact on the chemical evolution of these COMs. Instead, what matters is the increase in density and temperature as the collapse proceeds.

\section{Discussion} \label{sec:discussion}

\subsection{Evolution of the abundance profile}\label{sec:disc:profiles}

\begin{figure}
    \centering
    \includegraphics[width=0.49\textwidth, trim=0cm 0cm 0cm 0cm,clip]{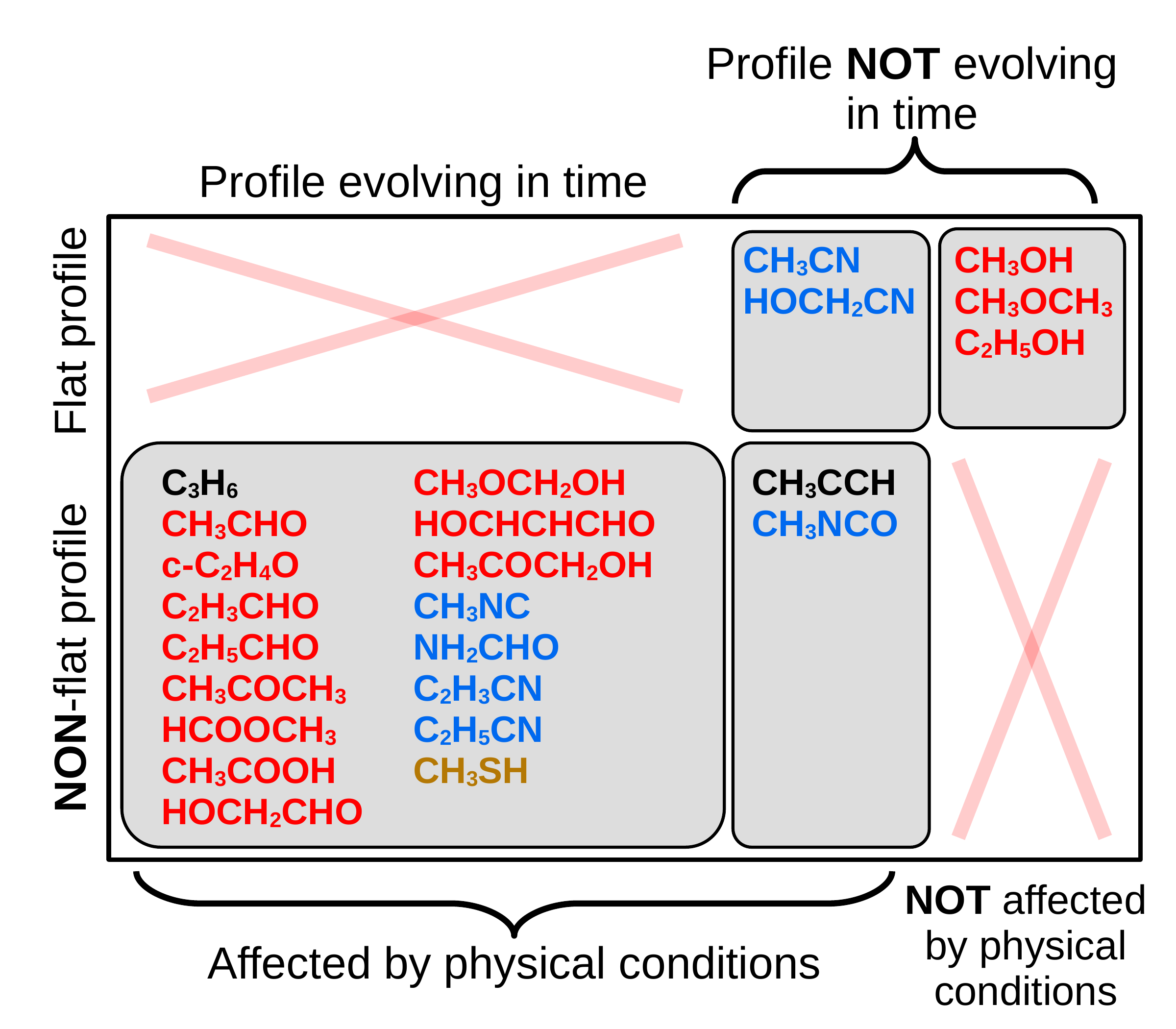}
    \caption{Molecules classified by the dependence of their abundance profile in the midplane of the disk. Top and bottom: Flat and nonflat abundance profiles, respectively. Left and right: Abundance profile evolving in time and not evolving in time, respectively, and/or affected or not affected by physical conditions. A species is considered unaffected if the radial profiles are within a factor of 2 from each other. There are no species whose profile evolves in time that display a flat profile in all conditions, and no species with a nonflat profile that are unaffected by physical conditions.}
    \label{fig:summary_affected}
\end{figure}

Our study shows that the abundance profile of complex organic molecules in the midplane of the disk can evolve from the early Class 0 to the late Class I. The chemical composition of a protoplanet can therefore be heavily impacted by the timing, location, and physical conditions of its formation. Figure \ref{fig:summary_affected} summarizes how the abundance profiles of COMs are affected by those parameters.

Only three species have flat abundance profiles that are  impacted neither by time nor by physical conditions: CH$_3$OH, CH$_3$OCH$_3$, and C$_2$H$_5$OH. Several other studies also find flat profiles of those species in Class 0 to Class II disks \citep[][although CH$_3$OCH$_3$ may be depleted in the inner $\sim 15$ au]{Drozdovskaya2016,Coutens2020,Navarro2024}. The abundance they find is also similar to ours, with differences less than a factor of a few \citep[see also][]{Walsh2014}. However, \citet{Drozdovskaya2016} determine that the abundance of CH$_3$OCH$_3$ and C$_2$H$_5$OH vary within one order of magnitude when assuming different physical conditions in the disk, due to different temperature profiles, which is more than in our case (less than a factor of 2). We identify two other species that also have flat abundance profiles that do not evolve in time, but that are affected by physical conditions: CH$_3$CN and HOCH$_2$CN. Notably, \citet{Legal2019} showed that the formation of CH$_3$CN could also be affected by the C/O ratio.
We do not find species that display a flat abundance profile and that evolve in time. Therefore, under given physical conditions, the initial relative abundances of CH$_3$OH, CH$_3$OCH$_3$, C$_2$H$_5$OH, CH$_3$CN, and HOCH$_2$CN in a protoplanet could be relatively independent of the location and time of its formation, at least before the Class II phase. In the inner few au of the disk (a region that is not considered here), UV dissociation, could however significantly impact COM abundances, particularly by depleting CH$_3$OH \citep{Nazari2022,Suzuki2024}.

By opposition, all other COMs are highly dependent on the radial distance, the age of the disk and the global physical conditions, except CH$_3$CCH and CH$_3$NCO that do not seem sensitive to the age of the disk. Several of those species vary by several orders of magnitude across the disk, in particular C$_3$H$_6$, CH$_3$CHO, C$_2$H$_3$CHO, C$_2$H$_5$CHO, HOCH$_2$CHO, NH$_2$CHO, and CH$_3$SH that become severely depleted in the inner $\sim 30$ au for disk ages $>10$ kyr. This behavior for CH$_3$CHO is in agreement with other models, as \citet{Coutens2020} obtain a flat density profile in a young $\sim 8$ kyr disk, while \citet{Drozdovskaya2016} observe a depletion in the inner $\sim 20 - 30$ au in their older Class II disks. However, HOCH$_2$CHO is not depleted in the inner $\sim 30$ au in one of the two disk models of \citet{Drozdovskaya2016}. We find in addition that a high cosmic-ray ionization rate causes the depletion of the same species between $\sim 25$ and $\sim 40$ au as early as $10$ kyr. These results show that those seven COMs are extremely sensitive to both spatial and temporal variations in physical conditions, suggesting that their radial distributions may serve as tracers of disk evolution and ionization.

\subsection{Inheritance}

All disk profiles contain all 26 COMs in significant abundance (on average at least $10^{-12}$, up to $\sim 10^{-5}$), which is not the case in the prestellar phase. While some species are already mostly formed during the prestellar phase, many are formed in majority during the collapse or in the disk. 

The formation of a given species does not necessarily depend on only one reaction pathway. For example, C$_2$H$_5$OH is formed in the cold prestellar phase in the grain mantle or on its surface from C+CH$_3$OH or C+CO followed by successive hydrogenations. Conversely, it primarily forms in warmer regions via diffusion with CH$_3$+CH$_2$OH in the grain mantle or on the grain surface. Other species, notably NH$_2$CHO, CH$_3$NCO, C$_2$H$_5$CN, and HOCH$_2$CN, also follow different formation paths in cold and hot environments. Those species are therefore formed both in the prestellar phase and the collapse through different paths, whose relative contributions vary with the physical conditions. 

Table \ref{tab:summary_formation} summarizes at which stage each species is formed, for each simulation at $t=\tff+10$~kyr. The label ``Early'' indicates species whose initial abundance is at least 50\% of the average disk abundance, while the label ``Late'' indicates species that have initial abundances less than 10\% of the disk average (so at least 90\% formed during the collapse or the disk). Species under the label ``Both'' are an intermediate case. They are formed in their majority (at least 50\%) during the collapse or in the disk, but are still significantly present during the prestellar phase (at least 10\%).

The fiducial case and the higher cloud mass yield similar results concerning inheritance, except for HOCH$_2$CN that is more significantly inherited at higher mass. Most of the lightest COMs are formed predominantly or significantly in the prestellar phase: CH$_3$CCH, C$_3$H$_6$, CH$_3$OH, CH$_3$CHO, CH$_3$OCH$_3$, C$_2$H$_5$OH, CH$_3$CN, CH$_3$NC, C$_2$H$_3$CN, and CH$_3$SH. Conversely, all the heaviest COMs are formed almost exclusively in the collapse or in the disk.

The results from the MHD simulations of \citet{Hincelin2013} and \citet{Coutens2020} also point toward an early formation of CH$_3$CCH, CH$_3$OH, CH$_3$CN, and CH$_3$SH, and a late formation of HCOOCH$_3$. However, contrarily to us, they find that CH$_3$CHO and CH$_3$OCH$_3$ form during the collapse. In \citet{Coutens2020}, this difference does not arise from the final abundances, which are similar to ours ($>10^{-7}$), but from the lower amount of molecules produced during the prestellar phase. In \citet{Hincelin2013} however, CH$_3$CHO is present in the disk in much lower quantity ($<10^{-8}$). \citet{Coutens2020} also find that different prestellar phase conditions may trigger an early formation of CH$_3$COCH$_3$ and NH$_2$CHO, which only happens for a higher molecular cloud temperature in our case.

A higher prestellar phase density reduces the importance of early formation of CH$_3$OH and C$_2$H$_5$OH, but increases those of N-bearing COMs, particularly CH$_3$NC, C$_2$H$_3$CN, and C$_2$H$_5$CN.
A higher molecular cloud temperature promotes an early formation of COMs, with all but five species being formed significantly during the prestellar phase: HCOOCH$_3$, HOCH$_2$CHO, HOC$_2$H$_4$OH, HOCHCHCHO, and CH$_3$COCH$_2$OH. Those COMs are among the heaviest considered here. Their initial abundance is negligible ($< 10^{-14}$) in all simulations compared to their average disk abundances ($> 10^{-10}$). Concerning the Early species, the quantity of molecules destroyed during the collapse amounts to less than half of their initial reservoir, which is therefore preserved in its majority from the prestellar phase to the disk. The only exception is again CH$_3$OH for which the amount destroyed during the collapse is equivalent to its initial reservoir.
A higher cosmic-ray ionization rate conversely diminishes the number of early formed species, particularly CH$_3$OCH$_3$, C$_2$H$_5$OH, and C$_2$H$_3$CN. The initial abundance of those three species are lower than in the fiducial case, while the disk averages are extremely close for CH$_3$OCH$_3$, C$_2$H$_5$OH, and higher (by a factor of $\sim 3$) for C$_2$H$_3$CN.
A shorter prestellar phase duration almost completely suppresses the early formation of COMs, which are all formed in negligible quantities compared to the subsequent collapse.
In summary, the inheritance is completely negligible for short durations of the prestellar phase, substantial for high molecular cloud temperatures, and significant only for the lightest COMs in the other cases.

Inheritance is often deemed important in the disk midplane due to its lower temperature and the shielding from UV photons that may otherwise dissociate or reprocess molecules \citep{Visser2009,Nazari2022}. Our results confirm that the disk midplane largely retains its reservoir of the simplest COMs from the prestellar phase, except when this phase is too short for significant COM formation. In this case however, most COMs that are formed early in the other simulations, namely C$_3$H$_6$, CH$_3$OH, CH$_3$CHO, CH$_3$OCH$_3$, C$_2$H$_5$OH, CH$_3$CN, CH$_3$NC, C$_2$H$_3$CN, and CH$_3$SH, form rapidly in $<50$ kyr at the beginning of the collapse. The minimal reprocessing after the ignition of the central object and the disk entry makes their chemical evolution close to an inheritance scenario. Notably, the inheritance of CH$_3$OH from the prestellar phase to the disk is widely supported by both models \citep{Yoneda2016,Drozdovskaya2016,Coutens2020} and observations \citep{Booth2021,Vandermarel2021,Booth2023,Booth2025}.

The outer disk ($\sim 10-20$ au near the outer edge) is also considered to be a key region for inheritance, as it is continuously replenished in new material coming from the envelope \citep{Visser2009,Eistrup2016}. Among the species significantly inherited from the prestellar phase, the abundance profiles of CH$_3$CHO may be a sign of this phenomenon. Its abundance at $r>40$~au is close to its initial value, but declines in the inner disk for $t\gtrsim \tff + 50$~kyr. As mentioned in Sect. \ref{sec:disc:profiles} other species exhibit similar profiles. This contrast between the inner and outer disk may therefore originate from this inheritance, although species other than CH$_3$CHO are mostly produced during the collapse, between the ignition of the central object and their entry in the disk.

\def\sl{\color{red} Late}
\def\se{\color{blue} Early}
\def\sm{\color{violet} Both}

\begin{table*}
  \caption{Stages at which COMs in disks are dominantly formed.}
  \label{tab:summary_formation}
\centering
\begin{tabular}{lllllll}
\hline\hline
        & Fiducial & Higher & Higher     & Higher      & Higher           & Shorter \\
Species & case     & cloud  & prestellar & mol. cloud  & cosmic-ray       & prestellar \\
        &          & mass   & density    & temperature & ionization rate  & duration\\
\hline
 CH$_3$CCH & \sm & \sm & \sm & \sm & \sm & \sl\\
 C$_3$H$_6$ & \sm & \sm & \sm & \sm & \sm & \sl\\
 \hline
 CH$_3$OH &  \se & \se & \sm & \se & \sm & \sl\\
 CH$_3$CHO &  \sm & \sm & \sm & \se & \sm & \sl\\
 c-C$_2$H$_4$O &  \sl & \sl & \sl & \se & \sl & \sl\\
 CH$_3$OCH$_3$ &  \sm & \sm & \sm & \se & \sl & \sl\\
 C$_2$H$_5$OH &  \se & \se & \sm & \se & \sl & \sl\\
 C$_2$H$_3$CHO &  \sl & \sl & \sl & \sm & \sl & \sl\\
 C$_2$H$_5$CHO &  \sl & \sl & \sl & \se & \sl & \sl\\
 CH$_3$COCH$_3$ &  \sl & \sl & \sl & \se & \sl & \sl\\
 HCOOCH$_3$ &  \sl & \sl & \sl & \sl & \sl & \sl\\
 CH$_3$COOH &  \sl & \sl & \sl & \se & \sl & \sl\\
 HOCH$_2$CHO &  \sl & \sl & \sl & \sl & \sl & \sl\\
 C$_2$H$_5$OCH$_3$ &  \sl & \sl & \sl & \se & \sl & \sl\\
 CH$_3$OCH$_2$OH &  \sl & \sl & \sl & \sm & \sl & \sl\\
 HOC$_2$H$_4$OH &  \sl & \sl & \sl & \sl & \sl & \sl\\
 HOCHCHCHO &  \sl & \sl & \sl & \sl & \sl & \sl\\
 CH$_3$COCH$_2$OH &  \sl & \sl & \sl & \sl & \sl & \sl\\
 \hline
 CH$_3$CN &  \sm & \sm & \sm & \se & \sm & \sl\\
 CH$_3$NC &  \sm & \sm & \se & \se & \sm & \sl\\
 NH$_2$CHO &  \sl & \sl & \sl & \sm & \sl & \sl\\
 C$_2$H$_3$CN &  \sm & \sm & \se & \se & \sl & \sl\\
 C$_2$H$_5$CN &  \sl & \sl & \sm & \sm & \sl & \sl\\
 CH$_3$NCO &  \sl & \sl & \sl & \sm & \sl & \sl\\
 HOCH$_2$CN &  \sl & \sm & \sm & \se & \sm & \sl\\
 \hline
 CH$_3$SH & \sm & \sm & \sm & \sm & \sm & \sl\\
\end{tabular} 
\tablefoot{The horizontal lines separate the hydrocarbons, and the O-bearing, the N-bearing, and the S-bearing COMs. Early: Species formed predominantly in the prestellar phase; Late: Species formed mostly during the collapse; Both: species formed mostly during the collapse, but at least 10\% formed in the prestellar phase.}
\end{table*}

\section{Conclusion} \label{sec:conclusion}

In this work we studied the evolution of 26 complex organic molecules during a protostellar collapse, up to 150~kyr after the formation of the circumstellar disk. We used the APE code to simulate the physical evolution of the gas in a protostellar envelope and a disk. With the Nautilus code, we then modeled the chemical evolution from the physical history of the gas. The initial abundances for those chemical calculations were obtained by simulating the evolution of atomic species in constant prestellar phase conditions. We focused our analysis on the chemical abundance of the COMs in the midplane of the disk at different times and for different physical conditions. We considered the total amount of molecules in the gas and solid phases. Below, we present our main results:

\begin{itemize}
    \item Two molecules are mainly formed during the prestellar phase and are inherited by the disk: CH$_3$OH and C$_2$H$_5$OH. Eight other species are also partially (between 10\% and 50\%) inherited by the disk from the prestellar phase: CH$_3$CCH, C$_3$H$_6$, CH$_3$CHO, CH$_3$OCH$_3$, CH$_3$CN, CH$_3$NC, C$_2$H$_3$CN, and CH$_3$SH. The 16 remaining molecules only form during the collapse or in the disk, mostly after the ignition of the central object.
    \item In the circumstellar disk, seven species show clear trends of total abundances (gas+ice+surface) increasing between disk ages of $10$ and $150$~kyr; specifically C$_3$H$_6$, C$_2$H$_3$CHO, C$_2$H$_5$CHO, CH$_3$COCH$_3$, C$_2$H$_3$CN, C$_2$H$_5$CN, and CH$_3$SH. Conversely, five species, heavy O-bearing COMs, have abundances decreasing with the disk age: HOCH$_2$CHO, CH$_3$OCH$_2$OH, HOC$_2$H$_4$OH, HOCHCHCHO, and CH$_3$COCH$_2$OH. The amplitude of those variations between disk ages of 10~kyr and 150~kyr can reach one order of magnitude. The average and median abundance of all other species do not vary significantly, except HCOOCH$_3$ whose average abundance varies by less than a factor of 2, but whose median abundance decreases by two orders of magnitude.
    \item Several species become severely depleted in the inner 30 au for disks older than 10~kyr: C$_3$H$_6$, CH$_3$CHO, C$_2$H$_3$CHO, C$_2$H$_5$CHO, HOCH$_2$CHO, NH$_2$CHO, and CH$_3$SH.

\end{itemize}

We also studied the sensitivity of COMs to the physical parameters: the density of the prestellar phase, the temperature of the molecular cloud, the prestellar phase duration, the cosmic-ray ionization rate, and the mass of the cloud (which affects the envelope density and the free-fall time). 
\begin{itemize}
    \item Changing the parameters generally leads to variations of up to one order of magnitude of the abundance profile. 
    \item A higher molecular cloud temperature promotes an early formation of COMs. Particularly, the reservoir of several molecules is fully formed during the prestellar phase under those conditions: c-C$_2$H$_4$O, CH$_3$OCH$_3$, C$_2$H$_5$OH, C$_2$H$_5$OCH$_3$, CH$_3$NC, C$_2$H$_3$CN, and HOCH$_2$CN. This reservoir is preserved in its majority during the collapse.
    \item A higher cosmic-ray ionization rate causes the depletion of seven species between $r\approx 25$~au and $r\approx 40$~au: C$_3$H$_6$, CH$_3$CHO, C$_2$H$_3$CHO, C$_2$H$_5$CHO, HOCH$_2$CHO, NH$_2$CHO, and CH$_3$SH. Those species are the same that are depleted in the inner disk at later times for the fiducial physical conditions.
    \item Increasing the mass of the protostellar cloud, and thus the duration of the collapse, does not significantly affect the abundance profiles in the disk midplane for most species. Only C$_2$H$_3$CHO, HCOOCH$_3$, CH$_3$COCH$_2$OH, and C$_2$H$_5$CN display profile features that are linked to the temperature, and are pushed back a few astronomical units due to the higher temperature of the central object.
    \item The abundance profile of most considered species is non-uniform, sensitive to the physical conditions, and sensitive to the age of the disk. The most notable exceptions are CH$_3$OH, CH$_3$OCH$_3$, and C$_2$H$_5$OH, whose total abundances are uniform in the midplane, and independent of the disk age and the physical conditions.
    \item The lightest COMs are significantly inherited in the disk from the prestellar phase for the fiducial case, and for the scenarios with a higher cloud mass, a higher prestellar density, and a higher cosmic-ray ionization rate. This notably concerns CH$_3$CCH, C$_3$H$_6$, CH$_3$OH, CH$_3$CHO, CH$_3$CN, CH$_3$NC, and CH$_3$SH. A higher molecular cloud temperature substantially increases the proportion of inherited material for all species but five heavy O-bearing COMs: HCOOCH$_3$, HOCH$_2$CHO, HOC$_2$H$_4$OH, HOCHCHCHO, and CH$_3$COCH$_2$OH. Conversely, a shorter prestellar phase duration inhibits the formation of COMs at this stage. Instead, the light COMs that would otherwise be inherited in the fiducial case form in the envelope at the beginning of the collapse.    
\end{itemize}

Our results highlight the key role of physical conditions in shaping the abundance of COMs and their inheritance in the midplane of disks, the primary site of planet formation. Future studies incorporating additional effects, such as shocks at the disk entry, will be essential to refining our understanding of the balance between inheritance and in situ chemistry in disk evolution.

\begin{acknowledgements}
 We thank the anonymous referee for their insightful comments that helped us improve the clarity and thoroughness of this paper. This study is part of a project that has received funding from the European Research Council (ERC) under the European Union’s Horizon 2020 research and innovation program (Grant agreement No. 949278, Chemtrip).
\end{acknowledgements}

\bibliographystyle{aa}
\bibliography{MaBiblio}

\appendix

\onecolumn
\section{Disk abundance profiles}\label{ann:profiles}

\begin{figure*}[h!]
    \centering
    \includegraphics[width=0.85\textwidth, trim=0cm 0cm 0cm 0cm]{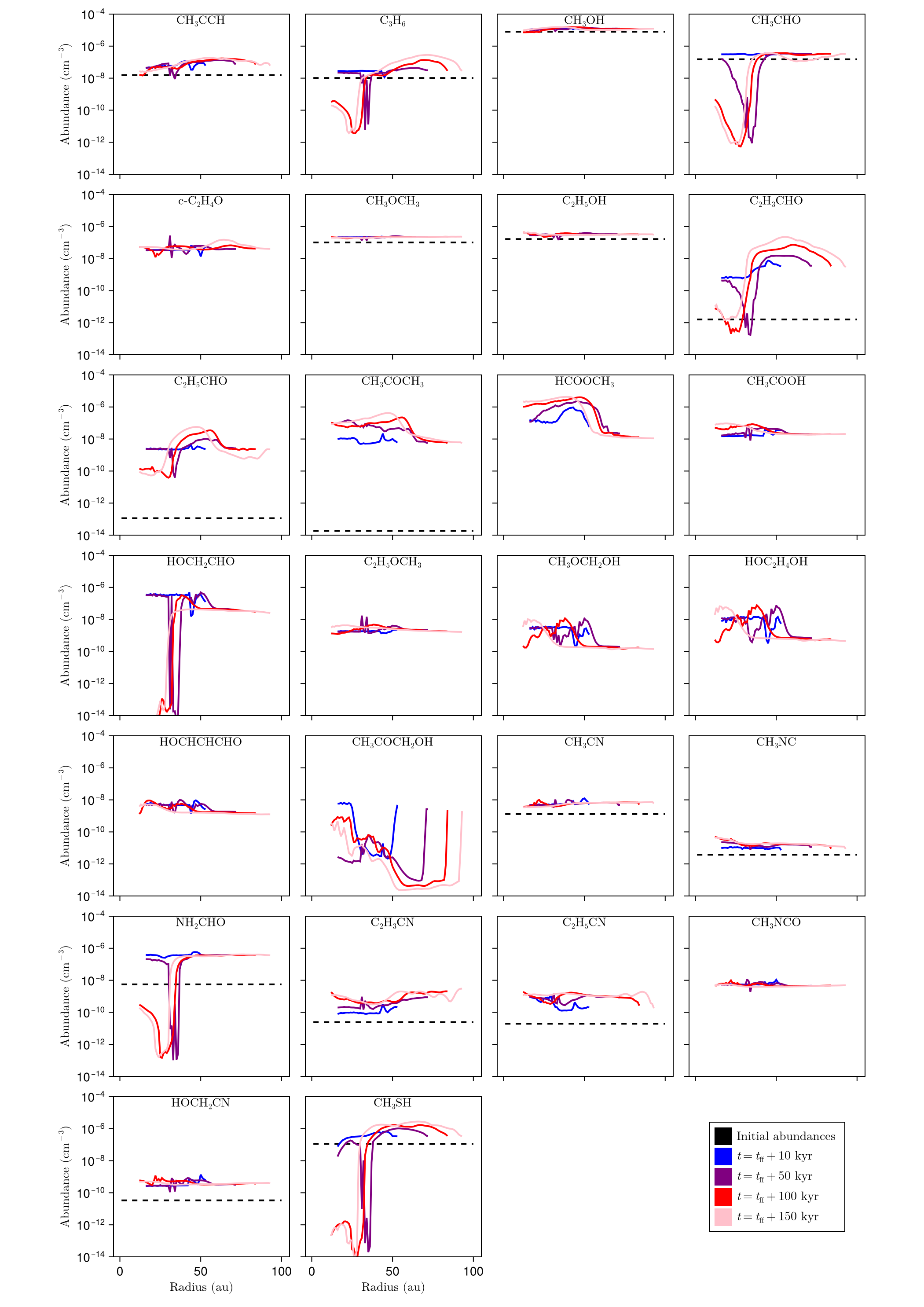}
    \caption{Profiles of total abundance (ice + gas) in the midplane of the disk of the COMs studied in this work. The colored lines represent different times after the formation of the central object: 10 kyr (blue), 50 kyr (purple), 100 kyr (red), and 150 kyr (pink). The dashed black line represents the initial abundance of the species (initial abundances below $10^{-14}$ are not displayed).}
    \label{fig:grid_times}
\end{figure*}

\begin{figure*}[h!]
    \centering
    \includegraphics[width=0.85\textwidth, trim=0cm 0cm 0cm 0cm]{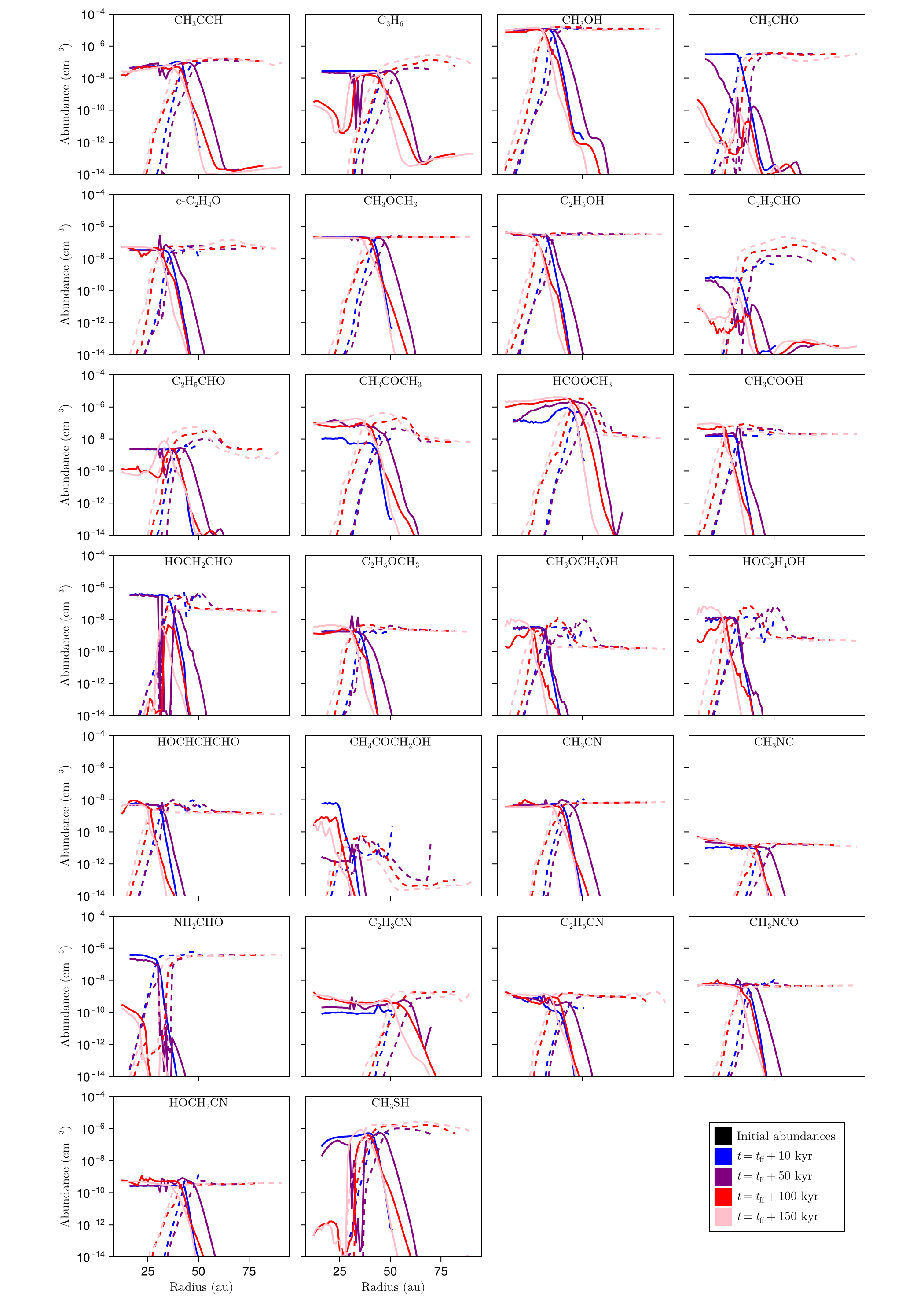}
    \caption{Same as Fig. \ref{fig:grid_times}, but separating the gas phase (solid lines) and ice phase (dashed lines).}
    \label{fig:grid_times_phases}
\end{figure*}

\begin{figure*}[h!]
    \centering
    \includegraphics[width=0.85\textwidth, trim=0cm 0cm 0cm 0cm]{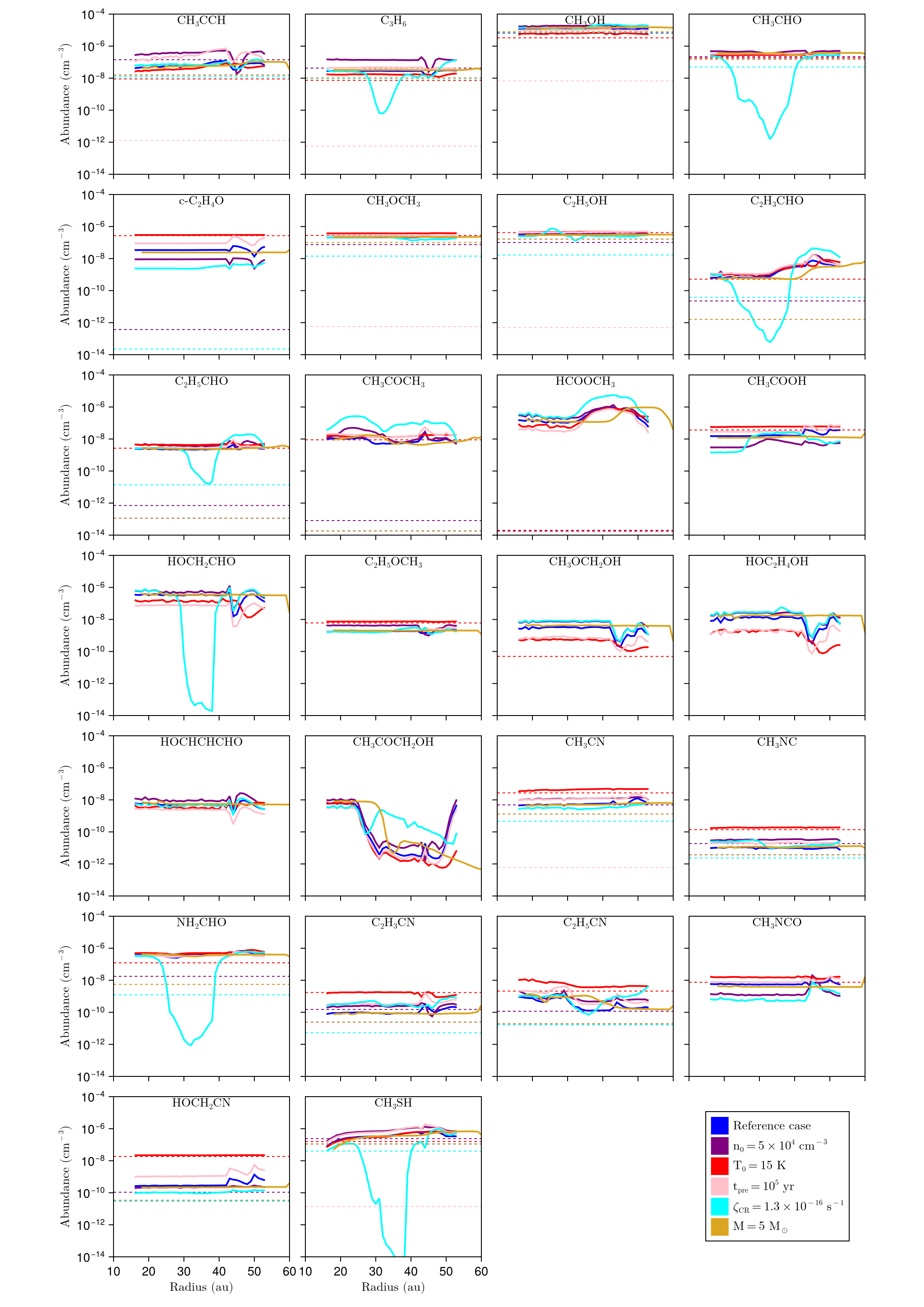}
    \caption{Profiles of total abundance (ice + gas) in the midplane of the disk of the COMs studied in this work. The colored lines represent different physical conditions. The dashed lines represent the corresponding initial abundances.}
    \label{fig:grid_preabund}
\end{figure*}

\begin{figure*}[h!]
    \centering
    \includegraphics[width=0.85\textwidth, trim=0cm 0cm 0cm 0cm]{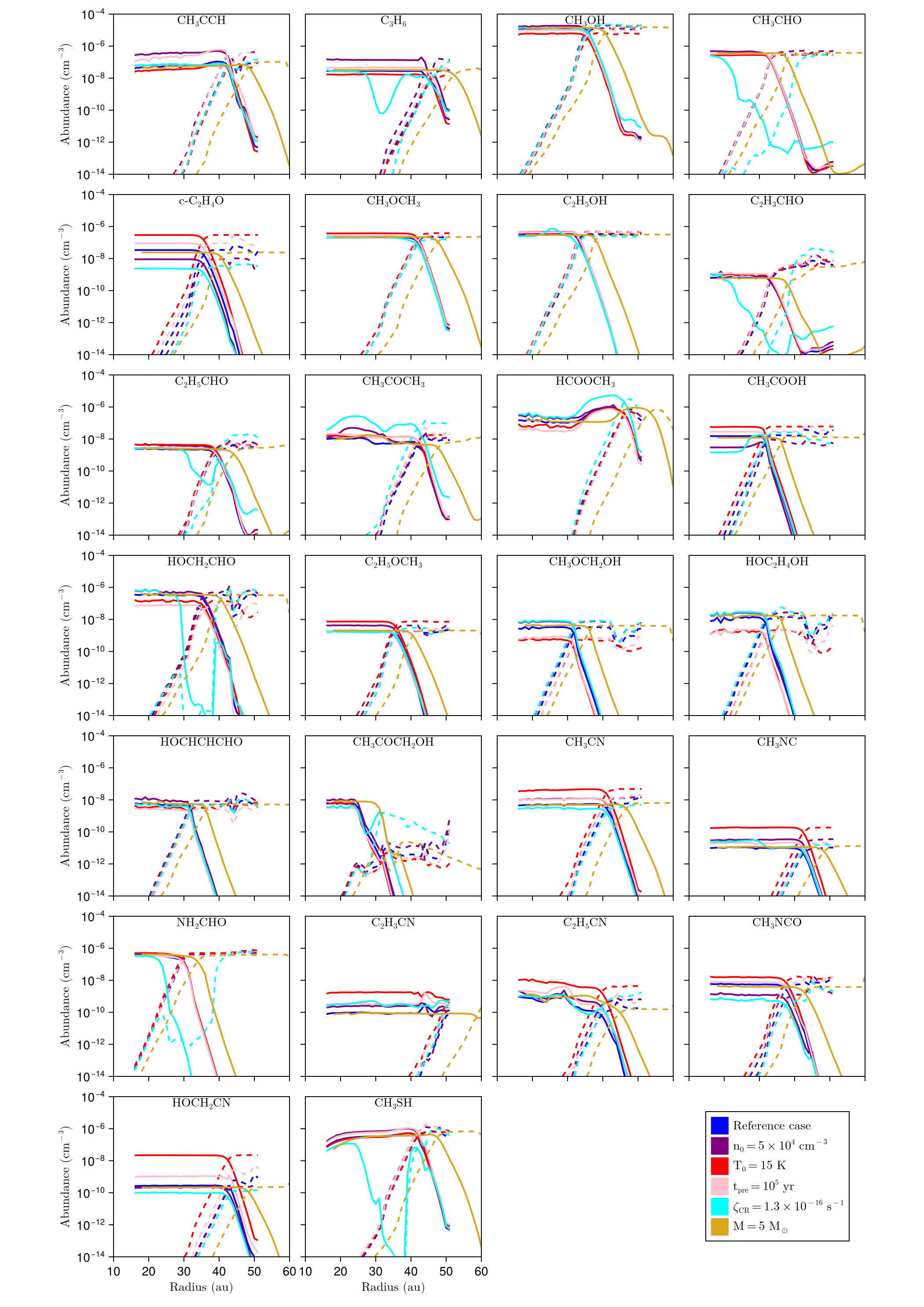}
    \caption{Same as Fig. \ref{fig:grid_preabund}, but separating the gas phase (solid lines) and ice phase (dashed lines).}
    \label{fig:grid_preabund_phases}
\end{figure*}

\end{document}